\def\lya{Ly$\alpha$}

\def\ha{H$\alpha$}
\def\hb{H$\beta$}
\def\oii{[O~II] $\lambda$3727}
\def\oiii{[O~III] $\lambda$5007}
\def\civ{C~IV~$\lambda$1550}

\def\exp#1{\times 10^{#1}}

\def\flux{\rm erg\ s^{-1} cm^{-2}}

\def\kms{{\rm km\ s^{-1}}}
\def\kmsmpc{{\rm km\ s^{-1}Mpc^{-1}}}

\def\msun{\rm M_{\odot}}
\def\sfr{\rm M_{\odot}\ yr^{-1}}

\def\deg{^{\circ}}
\def\etal{{\it et al.}}
\def\eg{{\it e.g.}}

\def\h0{H_0}
\def\q0{q_0}
\def\sles{\lower2pt\hbox{$\buildrel {\scriptstyle <}
   \over {\scriptstyle\sim}$}}
\def\sgreat{\lower2pt\hbox{$\buildrel {\scriptstyle >}
   \over {\scriptstyle\sim}$}}
\def\decsec{\hbox{$^{\prime\prime}\hskip-3pt .$}}
\def\sec{$^{\prime\prime}$}

\def\ref#1\par{\noindent \hangindent=4em \hangafter=1 #1 \par}  
\def\h100{h_{\rm 100}}
\def\mpc{{\rm Mpc}}

\documentstyle[12pt,aaspp4]{article}
\received{~}
\accepted{~}
\slugcomment{To appear in Ap.J.}

\begin{document}

\title{Keck Spectroscopy of Redshift $z\sim3$ Galaxies in the Hubble
Deep Field \footnote[1]{Based on observations obtained at the
W. M. Keck Observatory, which is operated jointly by the University of
California and the California Institute of Technology, and with the
NASA/ESA Hubble Space Telescope, which is operated by AURA, Inc.,
under contract with NASA.}}

\author{James D. Lowenthal$^{2}$\footnotetext[2]{Hubble Fellow},
David C. Koo, Rafael Guzm\'an, Jes\'us Gallego} 

\author{Andrew C. Phillips, S. M. Faber, Nicole P. Vogt, Garth
D. Illingworth} 

\and

\author{Caryl Gronwall}

\affil{UCO/Lick Observatory and Board of Astronomy and Astrophysics,
University of California, Santa Cruz, CA 95064.}
\affil{email: james, koo, rguzman, jgm, phillips, faber, nicole, gdi,
caryl@ucolick.org}

\begin{abstract} 

We have obtained spectra with the 10-m Keck telescope of a sample of
24 galaxies having colors consistent with star-forming galaxies at
redshifts $2\ \sles z\ \sles\ 4.5$ in the Hubble Deep Field (HDF).
Eleven of these galaxies are confirmed to be at high redshift ($z_{\rm
med}=3.0$), one is at $z=0.5$, and the other 12 have uncertain
redshifts but have spectra consistent with their being at $z>2$.  The
spectra of the confirmed high-redshift galaxies show a diversity of
features, including weak \lya\ emission, strong \lya\ breaks or damped
\lya\ absorption profiles, and the stellar and interstellar rest-UV
absorption lines common to local starburst galaxies and high-redshift
star-forming galaxies reported recently by others.  The narrow
profiles and low equivalent widths of C~IV, Si~IV, and N~V absorption
lines may imply low stellar metallicities.  Combined with the 5
high-redshift galaxies in the HDF previously confirmed with Keck
spectra by Steidel \etal\ (1996b), the 16 confirmed sources yield a
comoving volume density of $n \geq 2.5\exp{-4} \ h_{50}^3\ \mpc^{-3}$
for $q_0=0.05$, or $n \geq 1.2\exp{-3} \ h_{50}^3\ \mpc^{-3}$ for
$q_0=0.5$.  These densities are $3-4$ times higher than the recent
estimates of Steidel \etal\ (1996a) based on ground-based photometry
with slightly brighter limits, and are comparable to estimates of the
local volume density of galaxies brighter than L$^*$.  The
high-redshift density measurement is only a lower limit, and could be
almost three times higher still if all 29 of the unconfirmed
candidates in our original sample, including those not observed, are
indeed also at high redshift.  The galaxies are small but luminous,
with half-light radii $1.8 < r_{1/2} < 6.5\ \ h_{50}^{-1}$ kpc and
absolute magnitudes $-21.5 > M_B > -23$.  The HST images show a wide
range of morphologies, including several with very close, small knots
of emission embedded in wispy extended structures.  Using rest-frame
UV continuum fluxes with no dust correction, we calculate star
formation rates in the range $7 - 24$ or $3 - 9\ h_{50}^{-2}\ \sfr$
for $\q0=0.05$ and $\q0=0.5$, respectively.  These rates overlap those
for local spiral and H~II galaxies today, although they could be more
than twice as high if dust extinction in the UV is significant.  If
the objects at $z=3$ were simply to fade by 5 magnitudes (assuming a
$10^7$ yr burst and passive evolution) without mergers in the 14 Gyr
between then and now (for $q_0=0.05, h_{50}=1.0$), they would resemble
average dwarf elliptical/spheroidal galaxies in both luminosity and
size.  However, the variety of morphologies and the high number
density of $z=3$ galaxies in the HDF suggest that they represent a
range of physical processes and stages of galaxy formation and
evolution, rather than any one class of object, such as massive
ellipticals.  A key issue remains the measurement of masses.  These
high-redshift objects are likely to be the low-mass, starbursting
building blocks of more massive galaxies seen today.

\end{abstract}

\keywords{Galaxies: Redshifts -- Galaxies: Evolution -- Galaxies:
Formation -- Galaxies: Starburst -- Cosmology}

\section{Introduction}

Finding and studying populations of high-redshift galaxies is one of
the major goals of contemporary observational cosmology (Bahcall
1991).  By directly observing galaxies at lookback times corresponding
to 90\% of a Hubble time or more ($z \geq 3$ for $q_0 \geq 0.05$), we
can empirically constrain models of the formation and evolution of
galaxies.

It has been known for some time that broad-band colors can provide an
excellent means of estimating galaxy redshifts (e.g. Koo 1985; Lilly,
Cowie, \& Gardner 1991; Connolly \etal\ 1995), but only recently, with
the advent of 10-m class telescopes, have relatively normal galaxies
at $z>2$ become accessible for spectroscopic confirmation.  Because
the spectra of cosmologically distant, actively star-forming galaxies
show strong breaks at redshifted \lya\ and the Lyman continuum ---
imposed by internal interstellar absorption, intervening HI clouds,
the diffuse intergalactic medium, and other galaxies (\eg, Madau 1995)
--- they can be recognized via their generally blue colors coupled
with a dramatic dropoff in $U$ or $B$ (``$U$-'' or ``$B$-dropouts''),
corresponding to the redshifted \lya\ or Lyman continuum break passing
through those bands.

Steidel \etal\ (1996a) have pioneered the technique of using deep
broad-band photometry to isolate such Lyman-break galaxies, obtaining
confirming spectra at the 10-m W. M. Keck Telescope and high-resolution
images with the 2.4-m {Hubble Space Telescope} ({\it HST}) (Giavalisco
\etal\ 1996).  They found that this newly-discovered population of
$z>3$ galaxies exhibit stellar and interstellar absorption
lines characteristic of local starburst galaxies, as well as occasional
\lya\ emission lines. The objects have small sizes ($1.5 <
r_{1/2}< 3 \ h_{50}^{-1}$ kpc), high luminosities $L > {\rm L}^*$ and
star formation rates 12 - 75 $ h_{50}^{-2}\ \sfr$ (for $\q0=0.05$).
For $R_{AB} < 25.0$ and redshifts 3.0 - 3.5, the volume density was
found to be 0.1 - 0.5 times that of $L\ >\ {\rm L}^*$ galaxies today.

The Hubble Deep Field (HDF; Williams \etal\ 1996) provides an
unprecedented opportunity for locating and studying such high-redshift
galaxies.  With a total of 150 orbits of HST time invested in the
F300W, F450W, F606W, and F814W WFPC2 filters, final sensitivities
exceeded $I_{814,AB}=28$ ($3\sigma$) in a single WFPC2 field with area
4.7 arcmin$^2$. The HDF includes well over 1000 galaxies down to those
levels.

In this paper, we present spectra obtained at the Keck telescope of a
sample of $U$- and $B$-dropout galaxies selected by color to have
$z\sim3$ in the HDF.  This sample, observed as part of the ``DEEP''
program,\footnote[3]{Deep Extragalactic Evolutionary Probe; see
http://www.ucolick.org/$\sim$deep/home.html for more information} is
similar to that observed by Steidel \etal\ (1996b) in the HDF.
However, it has yielded more than twice the number of confirmed
redshifts $z>2.2$ and pushes roughly one magnitude deeper, sampling
further down the luminosity function of distant star-forming galaxies,
and significantly increasing the volume density of known high-redshift
galaxies.

\section {The Sample}

Many groups (\eg, Koo 1985; Bruzual \& Charlot 1993; Madau 1995;
Gronwall \& Koo 1995) have spent considerable effort studying the
expected colors of high-redshift galaxies in detail, including through
extensive modeling using synthetic and observed galaxy spectra,
stellar population considerations, and radiative transfer effects, and
more recently through analysis of deep spectroscopic surveys.  For our
observations of galaxies in the HDF, given the constraints of field
size, observing time and the relative immaturity of the technique,
we chose a simple set of selection criteria that maximized the chances
of finding high-redshift galaxies, but were not fine-tuned to exclude
interlopers.

We used the photometric catalog of objects in the HDF prepared by the
HDF team using FOCAS on the Version 1 ``drizzled''\footnote[4]{The
drizzled images result from a flux-conserving technique of combining
individual WFPC2 images registered to sub-pixel accuracy.  All four
chips (WF and PC) are resampled to 0\decsec04 pixels.  See
http://www.stsci.edu/ftp/observer/hdf for more information.} dataset
to select our sample.  We examined the distributions of isophotal
colors and magnitudes and selected 39 $U$-dropout objects, i.e. blue
in ``$B-I$'' ($B_{450} - I_{814} <1.22$)\footnote[5]{Note that we are
using AB magnitudes throughout, transformed from ST magnitudes as
detailed in the HDF information posted on the World Wide Web: to
transform from ST magnitudes to AB magnitudes, add 1.31, 0.399,
-0.199, and -0.819 to $U_{300},\ B_{450},\ V_{606},\ {\rm and\ }
I_{814}$, respectively.} but extremely red in ``$U-B$'' ($U_{300} -
B_{450} > 1.41$), the colors expected for a blue, star-forming galaxy
at $2<z<3.5$ with a Lyman continuum break (\eg, Madau 1996).  Objects
selected were brighter than $(V_{606}+I_{814})/2 = 25.5$ to help
guarantee adequate signal-to-noise ratios (SNRs) in the spectra.
Allowing objects up to one magnitude fainter into the sample would
have roughly doubled the number of candidates.  (Note that many of the
candidate objects do not completely disappear in the F300W image --
the Lyman continuum break falls within the F300W filter for $1.2 < z <
2.6$ -- although we retain the terminology ``dropout.'')

We also selected 11 $B$-dropout candidates.  Such objects are blue in
``$V-I$'' ($V_{606} - I_{814} <0.37$) but extremely red in ``$B-V$''
($B_{450} - V_{606} > 0.85$), and also have $(V_{606}+I_{814})/2 <
25.5$.  They result because (i) cosmologically intervening material is
expected to cause not only a Lyman continuum break but also a
significant \lya\ break (up to 1 magnitude at $z=4$; Madau 1995); and
(ii) the Lyman continuum break passes into the F450W filter
for $z>3.4$.  Four of these 11 $B$-dropout objects were already
in the $U$-dropout sample; thus we had a total sample of 46 candidate
high-redshift galaxies in the HDF.  Five of these were observed by
Steidel \etal\ (1996b), so we eliminated them from our target list
(although we were able to observe object C4-09 in the same slitlet
used for two other objects), leaving a final sample of 41 new
candidates.  We verified that our selection recovered all but one of
Steidel \etal's 18 additional candidates.  Note that no morphological
or size criterion of any sort was applied.

FOCAS splits large, complex ``parent'' objects into smaller
``offspring''; for our candidates, both parent and offspring objects
often satisfied the selection criteria.  We chose almost exclusively
the parent object, which was generally brighter and larger than its
offspring.  We inspected the image of each candidate visually to
verify that this assignment matched the appearance (spatial location
and apparent magnitude) of the candidate without including other
objects that were obviously unrelated, \eg, bright stars.

In Fig. 1 we show a mosaic of the 25 objects that we observed (except
for one object -- hd2\underbar{~}1918\underbar{~}1912, which turned
out to be a Galactic star also observed by Steidel \etal\ 1996b),
excised in 10\sec$\times$10\sec\ boxes from the F814W HDF images.  The
sample exhibits a wide variety of morphologies.  While it can be
immediately seen from Fig. 1 that almost all the targets are small --
with most of the light contained in 3\sec\ and usually within 1\sec\
-- only a relatively small fraction (6 of 24, or 25\%) have simple,
symmetrical, isolated profiles.  The remaining three-quarters all
exhibit various degrees of sub-clumping; extremely close companions
with colors similar enough to imply physical connection; extended,
wispy structures; or combinations thereof.  (Note, however, that we
see at most one example -- hd2{\underbar{~}}1739{\underbar{~}}1258 --
of a ``chain galaxy'' similar to those reported by Cowie, Hu, \&
Songaila 1995).

\section {Observations}

We used the Low Resolution Imaging Spectrograph (LRIS; Oke \etal\
1995) on the 10-m Keck telescope to obtain spectra on the nights of
22-24 April 1996 UT.  Five multi-object slit masks typically covered
10 high-redshift candidates at a time in the HDF, plus another 20
objects in the ``flanking fields'' selected for different programs
(Guzm\'an \etal\ 1996; Phillips \etal\ 1996; Vogt \etal\ 1996).  Most
of the high-redshift candidates were observed with more than one mask.
Each slitlet was 1\decsec1 wide and at least 10\sec\ long.  The slits
were up to 20\sec\ long in those cases when multiple objects could be
covered by a single slit, which was then tilted to the appropriate
position angle.  We took care beforehand to derive accurate
astrometric solutions for the HDF and the flanking fields (see our
page on the World Wide Web for details), and saw no evidence in the
data for astrometric errors.  The seeing ranged from 0\decsec7 to
1\decsec1 FWHM with photometric conditions prevailing on all three
nights.

We chose to use two moderate-resolution gratings --- both 600 l/mm and
blazed at 5000 \AA\ (blue) and 7500 \AA\ (red) --- rather than one
low-resolution grating.  This choice helped in subtraction of bright
night sky emission lines and provided the best information possible on
emission and absorption line widths and profiles.  The slit and
grating combinations produced a spectral resolution of 4.5 \AA\ FWHM
at 5500 \AA, or about $240\ \kms$ ($\sigma = 100\ \kms$), for a source
that completely fills the slit.  However, for the objects discussed
here, which are almost all unresolved at ground-based seeing, the
resolution is determined by the seeing disk and not the slit.  In this
case, the resolution is $\sim 3.4$ \AA\ FWHM, or $\sigma = 185\ \kms$.
The spectral coverage was complete from about 4000 \AA\ to 9000 \AA,
the exact range depending on the location of the individual slitlets
on the slit masks.  Exposure times were $2\times1500$ sec per mask in
each of the two gratings, for a total of 6,000 sec (8 targets,
observed with only one mask) or 12,000 sec (17 targets, observed with
two masks) per object, evenly split between the two gratings..

The data were reduced and analyzed using IRAF tasks and some of our
own custom software.  The images were bias-subtracted, flat-fielded,
cosmic-ray cleaned, rectified using an analytical model to remove
geometrical optical distortion and to straighten the spectra
obtained through tilted slits, and sky-subtracted.  Finally the
two-dimensional images were co-added for each individual object and
optimally extracted to a one-dimensional format.  In practice, only
the blue spectra were useful, as the blue colors of our sample
galaxies meant that the continuum was often not visible in the red. 
Wavelength calibration was performed on the rebinned,
two-dimensional spectra by using the bright 5577 \AA\ O~I night sky
emission line as a zero-point reference; rms errors in the solutions
are $<0.2$ \AA.

Table 1 lists the name, coordinates, apparent magnitude, color, class
($U$- or $B$-dropout), and exposure time of all 25 objects.  In
those cases for which the FOCAS catalog gave $U_{300,AB}>28.0$, close
to the detection limit for a compact ($<1$\sec) source in the HDF, we
list a conservative lower limit to $U-B$ calculated for
$U_{300,AB}=28.0$.  In theory, the catalog contains useful information
on the SNR of each object that could be applied on a case-by-case
basis.  However, at the time of our sample selection the noise
properties of the HDF were not well characterized, so we chose the
simpler and more robust route.  In a few cases, measurement errors at
these faint magnitudes meant that only the parent object met the color
selection criteria and not the offspring, or vice versa; in these
cases the candidate was retained although the tabulated color might
lie slightly outside the selection zone.

\section {Redshifts and Spectral Characteristics}

We show in Fig. 2 one- and two-dimensional spectra for the 12 objects
that yielded secure redshifts.  Six of these exhibit single emission
lines that we identify as \lya; almost all show either a significant
\lya\ break or a damped \lya\ absorption profile.  In addition, many
of the spectra show the strong stellar and interstellar absorption
lines of O~I, Si~II, C~II, Al~II, and C~IV that are seen in other
recently-reported high-redshift galaxies (Steidel \etal\ 1996a; Ebbels
\etal\ 1996; Yee \etal\ 1996; Trager \etal\ 1996); in some cases,
these features are the only ones that indicate the redshift.  Only one
object, hd4\underbar{~}1229\underbar{~}1791, was found to be at $z <
2$.  Table 2 lists all twelve of the confirmed redshifts.

In an attempt to extract redshifts from those spectra with low SNR, we
cross-correlated each spectrum with a one-dimensional spectrum of NGC
1741, a nearby starburst/Wolf-Rayet (WR) galaxy, obtained by Conti,
Leitherer, \& Vacca (1996) with HST and the Goddard High Resolution
Spectrograph and kindly provided in digital form by C. Leitherer.  In
every case where there were obvious spectral features, the
cross-correlation function showed a strong and isolated peak.
However, in those cases without strong spectral features, the
cross-correlation generally produced unconvincing results that we took
only as a suggestion to be followed up with a visual search for
identifiable spectral features.  These uncertain values are shown in
column 2 of Table 3.

There are at least two objects that show evidence for multiple
absorption systems at different redshifts.  One of these,
hd4{\underbar{~}}1076{\underbar{~}}1847, is the brightest galaxy in
the sample but has not yet yielded a confirmed redshift.  This object
is located in an especially complex region, only 3\sec\ from object
C4-09 at $z=3.22$ (Steidel \etal\ 1996b), 7\sec\ from
hd4{\underbar{~}}1229{\underbar{~}}1791 at $z=0.483$ (the only
confirmed redshift $z<1$ in our study), 2\sec\ north of a galaxy at
$z=1.010$ observed by Cohen \etal\ (1996), and 1\decsec5 east of a
galaxy at $z=0.882$, also observed by Cohen \etal.  These latter two
objects are apparently responsible for
Mg~II~$\lambda\lambda~2796,2803$ absorption visible at those redshifts
in our spectrum, which is shown in Fig.~3.  The projected distance of
1.5\sec - 2\sec\ between hd4{\underbar{~}}1076{\underbar{~}}1847 and
the absorbing galaxies corresponds to 15 - 20 $h_{50}^{-1}$ kpc
($q_0=0.05$) at $z\sim1$, well within the range of typical absorption
cross-sections for QSO Mg~II absorption line clouds (Drinkwater,
Webster, \& Thomas 1993; Steidel 1995).  Galaxy
hd4{\underbar{~}}1076{\underbar{~}}1847 may itself be at $z=1.01$, of
course, although we tentatively assign $z=2.15$ based on possible
\civ\ (and maybe Al~II) absorption.

The other object showing evidence for multiple redshift absorption is
hd2{\underbar{~}}0698{\underbar{~}}1297, which has its own confirmed
redshift of $z=3.430$.  We find weak absorption lines plausibly
matching Si~II 1260 \AA\ and O~I 1303 \AA\ at $z=3.368$, the redshift
of hd2{\underbar{~}}0705{\underbar{~}}1366, only 3\sec, or $12\
h_{50}^{-1}$ kpc projected distance ($q_0=0.05$), away (although the
signal-to-noise ratio is low).  This is again comparable to
typical absorption cross-sections of Mg~II-selected galaxies at
moderate redshifts.

Tables 2 and 3 give the redshift obtained for every object in the
sample. The redshift quality $Q_z$ is given on a scale from 1 to 4
such that $Q_z=1$ means there is little hope of assigning a redshift
given the SNR in our data, 2 means real features are evident but the
redshift is not secure, 3 means the redshift is probable, and $Q_z=4$
means the redshift is definitely secure, with multiple spectral
features identified.

The median redshift for the 11 galaxies in our sample with $Q_z \geq
3$ and $z>2$ (``confirmed high redshift'') is $z_{\rm med}=3.0$;
restricting the sample to $Q_z = 4$ does not change the median.  Eight
of the confirmed high-redshift galaxies were in the $U$-dropout
sample ($z_{\rm med}=3.0$), seven were in the $B-$dropout sample
($z_{\rm med}=3.2$), and four were in both samples.  Meanwhile, the
fraction of $B$-dropouts with confirmed redshifts $z>2$ is 0.78
vs. 0.40 for $U$-dropouts.  This underscores the importance of the
$B-V/V-I$ color criterion's sensitivity, both to the \lya\ break at
$2.3 < z < 3.3$ and to the Lyman continuum break at $3.4 < z < 4.7$.

Apart from hd4\underbar{~}1229\underbar{~}1791 at $z=0.483$ (and
hd2\underbar{~}1918\underbar{~}1912, the Galactic star), there is no
evidence for redshifts lower than $z=1.3$ or higher than $z=5$; this
includes the 12 galaxies {\em not} in the confirmed sample (with the
possible exception of hd2\underbar{~}0853\underbar{~}0319, which
resembles hd4\underbar{~}1229\underbar{~}1791 in morphology and
color).  If any other galaxies in our sample had $z<1.3$, we would
expect to see features such as \oii\ or the 4000 \AA\ break.  If any
had $z>5$, no flux should be visible below 5500 \AA\ due to absorption
below the Lyman continuum break, while in fact flux is visible below
that wavelength in all cases.  Each galaxy's red spectrum was
inspected for additional emission lines or other spectral features
that might support or refute the redshift assigned on the basis of the
blue spectrum; no such features were found.

To investigate the galaxies' spectral properties further, we added
together the spectra of all the confirmed high-redshift objects to
produce a rest-frame average spectrum, shown in Fig. 4.  In addition
to the strong \lya\ continuum break and the stellar and interstellar
absorption lines mentioned above, the average spectrum shows
absorption due to Ly$\gamma$, Ly$\beta$, and Fe~II $\lambda1608$, and
weak emission corresponding to C~III] $\lambda1909$ (although we note
that the signal at the two ends of the average is dominated by only a
few objects, due to the different rest-frame spectral ranges covered
by the individual spectra).  The latter feature is commonly seen in
the spectra of active galactic nuclei (AGN), but it is also associated
with planetary nebulae.  The high-ionization lines of N~V, Si~IV,
and C~IV may hold some clues to the metallicity and stellar content of
high-redshift star-forming galaxies; this is discussed in \S 5 below.

Fig. 5 shows redshifts vs. $I_{814,AB}$ for the confirmed sample.
Also shown for comparison are (i) the faint magnitude limit of the
Canada France Redshift Survey (Lilly \etal\ 1996); (ii) three lines
corresponding to unevolved L$^*$, 10L$^*$, and 100L$^*$ based on
NGC4449, a blue star-forming galaxy; (iii) two galaxies that have been
interpreted as possible primeval galaxies, IRAS 10214+4724
(Rowan-Robinson \etal\ 1991) and CNOC cB58 (Yee \etal\ 1996), although
both those galaxies may be gravitationally lensed and thus magnified
by up to a factor of 10; (iv) an \ha-emitting galaxy at $z=2.5$ found
near a QSO absorption-line cloud (Malkan, Teplitz, \& McLean 1995,
1996); (v) the gravitationally lensed high-redshift galaxies
discovered by Trager
\etal\ (1996); and (vi) the five confirmed high-redshift objects in
the HDF observed by Steidel \etal\ (1996b).  Also included are two
serendipitous objects at $2<z<3$ in the HDF flanking fields that were
measured as part of a program to study compact objects (see Phillips
\etal\ 1996 for more details).

The DEEP sample probes significantly fainter than most previous
studies of high-redshift galaxies: the confirmed sample comprises
objects that are barely more luminous than L$^*$ galaxies today.  Note
too that the confirmed sample has a median $I_{814,AB}=25.0$,
compared to the median $I_{814,AB}=24.0$ for the 5 galaxies from
Steidel \etal\ (1996b), who were restricted by bad weather to only a
few hours total integration time.

It appeared from the uneven spatial distribution of high-redshift
candidates on the three WFPC2 chips that there was evidence for
clustering (as pointed out by Steidel \etal\ 1996b): 23 of the 46
original candidates are on WF2, 8 on WF3, and 20 on WF4.  However, the
redshift distribution shows that most of the apparent ``clustering''
on WF2 and WF4 is due to chance superposition rather than physical
association.  There is some hint of a group near $z=3$: three galaxies
within a circle of radius 85\sec\ have confirmed redshifts $z=2.980,
2.990, $ and 2.991 (including iw2{\underbar{~}}0547{\underbar{~}}0293,
one of the compact objects from the Flanking Field sample of Phillips
\etal\ 1996).  These values translate to $\sim\ 1$ Mpc and a
rest-frame velocity spread of $\sim\ 3300\ \kms$, comparable to that
in dense regions (clusters or walls) of galaxies at the current epoch,
and to another possible group in the HDF pointed out by Steidel \etal\
(1996b).

Several workers have used the observed colors of galaxies in the HDF
to assign ``photometric redshifts'' and thus attempt to identify
high-redshift candidates in a variation on the techniques used to
select the present sample (Clements \& Couch 1996; Lanzetta, Yahil,
\& Fern\'andez-Soto 1996; Madau \etal\ 1996; Mobasher
\etal\ 1996).  Column 7 of Table 2 indicates where those studies
overlap with our sample.  We have compared the redshift predictions of
Lanzetta \etal\ 1996 with our confirmed observed redshifts, including
the Galactic star and the galaxy at $z=0.483$, and find that there is
a rough correlation with an offset $z_{DEEP} - z_{Lanzetta} = 0.35$
and an rms dispersion of $0.29$ in $z$, not including one outlier at
$z>3$ that was predicted to have $z<0.25$.

In Fig. 6, we plot the colors of the eleven confirmed high-redshift
galaxies and compare them to the regions of color-color space chosen
by Madau \etal\ (1996) to select high-redshift candidates.  There are
five objects -- almost half our sample -- that do not satisfy those
criteria and yet are confirmed to have $z>2.9$.  The criteria chosen
by Madau \etal\ were designed to be efficient but also relatively free
from contamination by low-redshift galaxies.  Nevertheless, the
regions of color-color space from which we drew our sample seem
reasonably well-separated from the remainder of the HDF field galaxies
at the magnitudes accessible by spectroscopy at Keck, and our
contamination was limited to one Galactic star and one galaxy at
$z=0.483$.  These differences demonstrate that the techniques used by
Madau \etal\ and Lanzetta \etal\ are promising but not yet secure;
some additional experimentation with color selection and calibration
of the photometric redshift assignments using the present data will no
doubt help refine the techniques.

\section {Intrinsic Properties and Speculations on the Nature of $z=3$
Galaxies}

\subsection {Sizes, Luminosities, and Star Formation Rates}

What are the $z=3$ galaxies in the HDF, and how do they correspond to
galaxies in the nearby Universe today?  We have calculated absolute
luminosities $M_B$ and intrinsic half-light radii $r_{1/2}$ based on
measured redshifts, apparent magnitudes and sizes in the HDF images.
We assumed $k$-corrections derived for a young, non-evolving galaxy
with constant star formation, but including a minimal amount of dust
extinction ($E(B-V)=0.1$ with an SMC-type reddening curve; ``Class 1''
of Gronwall \& Koo 1995).  These parameters are listed in Table 2 and
plotted in Fig. 7, which also shows for comparison the sizes and
luminosities of a large sample of nearby galaxies with a wide variety
of Hubble types.  In the four cases where the source was clearly
separated into discrete multiple components in the HDF images, we show
the parameters for both the aggregate and individual components.

Fig. 7 reveals that the HDF galaxies at $z=3$ are intrinsically quite
small, with a median half-light radius $r_{1/2}=3.6\ h_{50}^{-1}$ kpc
and a range $1.7 < r_{1/2} < 7\ h_{50}^{-1}$ kpc (for $q_0 = 0.05$;
for $q_0 = 0.5$, values would be $\sim$ 40\% smaller).  This
corresponds to the sizes of large dwarf spheroidals, H~II galaxies,
and Compact Narrow Emission Line Galaxies (CNELGs; Koo \etal\ 1994;
Koo \etal\ 1995; Guzm\'an \etal\ 1996), and small ellipticals and
spiral bulges.  However, they are quite luminous, with median absolute
magnitude $M_B = -22.5 + 5\ {\rm log}\ h_{50}$, similar to those of
the brightest normal galaxies in the local Universe and $\sim 1.5$
magnitudes brighter than M* (Lin
\etal\ 1996).  It is also apparent that the $z=3$ galaxies overlap
with the bright end of the sequence defined by local HII galaxies and
distant CNELGs.

Following the prescriptions of Warren \etal\ (1995) and Steidel
\etal\ (1996a), which are based on the models of Leitherer, Robert,
\& Heckman (1995) and Kennicutt (1983), we have used rest-frame UV
luminosities to estimate star formation rates, assuming continuing
star formation and a Salpeter IMF with an 80 $\msun$ upper mass
cutoff.  SFRs for {\em total} objects (not subclumps) range from 7 to
25 $h_{50}^{-2} \sfr$ for $q_0=0.05$ (or 3 to 8 $\ h_{50}^{-2}
\sfr$ for $q_0=0.5$).  These values span the range found in late-type
spiral and HII galaxies today (Kennicutt 1983; Gallego \etal\ 1995;
Telles 1995).  They are somewhat lower than the star formation rates
reported by Steidel \etal\ (1996a,b; typically $SFR \sim 25
h_{50}^{-2} \sfr$ for $q_0=0.05$), which simply reflects the fainter
continuum levels reached in our sample.

Note that we have not applied any extinction correction to the rest-UV
fluxes observed in the F300W and F450W filters; including an SMC-type
extinction law with $E(B-V)=0.1$ will produce about 1 magnitude of
extinction at 1500 \AA\ (Gronwall \& Koo 1995; Bouchet \etal\ 1985),
which would make the star formation rates higher by more than a factor
of 2.

The star formation rates quoted above would produce, in the absence of
absorption, \lya\ emission lines with rest equivalent widths on the
order of 100 \AA\ (Charlot \& Fall 1993) and observed fluxes on the
order of $1\exp{-16} \flux$ (assuming Case B recombination and the
relations calculated by Kennicutt 1983).  These are well above the
detection thresholds of recent searches for \lya\ emission in ``blank
sky'' (Lowenthal \etal\ 1991; Thompson \& Djorgovski 1995).  The low
or undetected equivalent widths of \lya\ emission observed in all
objects in our sample (median for seven detections is $W_0 = 12$ \AA)
imply that there is significant internal absorption, presumably due to
small amounts of dust, and demonstrate why the blank sky searches have
been uniformly unsuccessful at detecting high-redshift galaxies.

\lya\ emission has also been especially elusive from damped \lya\
QSO absorption line clouds at high redshift, despite extensive
searches (see Lowenthal \etal\ 1995 for a summary, and Djorgovski
\etal\ 1996 for a recent detection).  The spectra in our confirmed
sample and in the somewhat brighter samples of Steidel \etal\ (1996
a,b) seem to show a damped \lya\ absorption profile, on average (see
Fig. 4), so we can assume that these galaxies have a sufficiently high
cross-section of neutral hydrogen to cause damped
\lya\ absorption in QSO spectra as well.  Thus the
Lyman-break-selected galaxies are probably a subset of the damped
\lya\ clouds.  The weak or absent \lya\ emission lines in our
current sample are entirely consistent with the upper limits and weak
detections reported from damped \lya\ cloud searches.

The six \lya\ emission lines have FWHMs ranging from 5 to 10
\AA, which, after subtracting quadratically the instrumental FWHM of
3.4 \AA\ and correcting to rest-frame values of velocity dispersion
$\sigma$, correspond to a range $100 < \sigma\ < 230\ \kms$
($\sigma_{\rm med} = 150\ \kms$), with an estimated uncertainty on the
order of 20\%.  The highest SNR \lya\ emission line (SNR $\sim 15$)
has the lowest measured velocity dispersion ($\sigma = 100\ \kms$) in
that range.  The implications and possible systematic effects are
discussed in the following section.

We may draw some inferences on the physical properties of
high-redshift galaxies based on the absorption lines as well.  The
Si~IV and C~IV absorption lines seen in the average spectrum in Fig. 4
are typically associated with hot stars and stellar winds, but they
also arise in the interstellar medium (ISM; \eg, York \etal\ 1990;
Leitherer \etal\ 1996).  The narrow profiles we observe in our average
spectrum certainly resemble those observed in a star cluster in NGC
1705 that has an age $> 10$ Myr and no O stars (Leitherer 1996; Meurer
\etal 1992) more than they do the deep, stellar wind-broadened
profiles seen in NGC 1741, the comparison starburst/WR galaxy shown in
Fig. 2.

Furthermore, Walborn \etal\ (1995) found that early O~V stars in the
Small Magellanic Cloud (SMC), which has low metallicity ($\rm [Fe/H] =
-0.65$; Russell \& Bessell 1989), generally exhibit significantly
weaker and narrower C~IV, N~IV, and N~V stellar wind absorption
profiles than their counterparts in the higher metallicity Large
Magellanic Cloud (LMC; $\rm [Fe/H] = -0.30$), although the
physical correlation is not well-established.  Presumably the
difference in line strengths is due to the lower terminal velocities
that a low-metallicity environment produces in a radiation-driven
wind.  The N~V, Si~IV, and C~IV absorption lines in our spectra of
$z\sim3$ galaxies resemble the lines in the spectra of SMC stars more
than those of the LMC stars.  If the absorption lines do indeed arise
in the atmospheres of hot stars and not the ISM, then those stars are
apparently low metallicity, consistent with an early stage of star
formation.

\subsection{Evolution, Masses, and Space Densities}

Fig. 7 may provide some insight into the eventual fate of the
high-redshift galaxies by indicating what will happen as they evolve
with cosmic time.  Under the assumptions that they do not undergo
significant merging and that their current bursts of star formation
fade in the future, they will simply move down in the $M_B-r_{1/2}$
diagram, retaining their current size.  Using updated versions of
Bruzual \& Charlot's (1993) spectral synthesis galaxy evolution code,
we calculated the evolution of a $10^7$ yr burst model assuming a
Salpeter IMF with stellar masses 0.1 - 125 $\msun$ and passive
evolution.  From $z=3$ to $z=0$, the expected amount of fading is
$\sim 5$ mag.  This would place the $z=3$ galaxies close to the region
occupied by dwarf spheroidal and irregular galaxies and spiral bulges
today, but not near normal massive ellipticals or total spiral
half-light radii.

If the bright cores of UV-luminous star-forming regions that we are
now observing reside in the centers of larger accumulations of older,
evolved populations of stars, then the apparent half-light radii will
grow somewhat as the core fades.  However, in H~II galaxies and
IR-selected starburst galaxies this effect typically amounts to only a
factor of two increase in $r_{1/2}$ (Lehnert \& Heckman 1996), still
failing to reconcile the objects we observe at $z=3$ with normal,
massive spiral and elliptical galaxies today.  Deep near-IR
photometric images, such as those recently obtained by the HDF team at
the Kitt Peak National Observatory 4-m telescope and those planned by
the NICMOS team with HST, will help address that issue by revealing
any massive, evolved stellar populations underlying the high-redshift
galaxies.  The possibility of growth in $r_{1/2}$ through mergers is
discussed in \S 5.3 below.

Steidel \etal\ (1996a,b) and Giavalisco \etal\ (1996) have suggested
that $z\sim3$ $U$-dropout galaxies, including those in the HDF, may
represent the cores of proto-spheroids that are forming stars at the
centers of deep, {\em massive} potential wells.  They cite as evidence
the apparently smooth, compact morphologies in their sample, the
agreement between the half-light radii of the high-redshift galaxies
and those of spheroids today, and the high equivalent widths of the
saturated interstellar UV absorption lines, which they tentatively
interpret as implying virial velocities on the order of 200 $\kms$,
consistent with massive galaxies.  The confirmed high-redshift
galaxies in the DEEP sample also show sizes similar to those of
spheroids today, where by spheroids we mean the dynamically relaxed,
pressure-supported portions of disk galaxies, including bulges and
halos, and of ellipticals.

However, our sample of confirmed (and also candidate) high-redshift
galaxies in the HDF is far from homogeneous morphologically,
suggesting instead that we are observing a diverse range of physical
processes or stages of formation and assemblage of galaxies.  As
proposed by Van Den Bergh \etal\ (1996), this discrepancy may be due
at least in part to the shallower depths compared to the HDF reached
in previous HST images, which would then tend to show only the
brightest, most compact, and possibly most uniform features.

Furthermore, as pointed out by Conti \etal\ (1996), the large
equivalent widths of saturated interstellar lines in such galaxies as
NGC 1741, the starburst galaxy whose spectrum closely resembles those
of our sample, cannot be used to measure masses with any certainty;
this is because the lines may be broadened by such processes as
supernova winds and shocks, rather than simply by virial motions.  In
fact, the average rest equivalent width of the unblended interstellar
absorption lines in NGC 1741 is 2 \AA; if the line widths reflect
virial motions, this translates to a velocity width $\sigma
\sim 200\ \kms$ or a gravitational mass $>10^{11}\ \msun$.  However, the gas
mass is only $M(HI)=10^{10}\ \msun$ (Williams, McMahon, \& Van Gorkom
1991) and the total mass is probably $M_{tot}\ll 10^{11}\ \msun$.
Such equivalent widths are typical of those in the high-redshift
sample, so caution should be exercised in deriving virial masses --
and broader cosmological conclusions (\eg, Mo \& Fukugita 1996) --
therefrom.

At the same time, the small widths of the six \lya\ emission lines we
observe ($100 < \sigma < 230\ \kms$) imply upper limits to
gravitational masses on the order of $10^{10} \msun$, small in
comparison to massive galaxies today.  This mass constraint may be
more robust than the absorption line equivalent widths, since most
radiative transfer effects act to broaden \lya, although \lya\
emission is certainly complex and may be asymmetrically absorbed or
originate in only a small fraction of the total star-forming area
(\eg, Auer 1968; Neufeld 1991; Charlot \& Fall 1993).  

One exception may be the case of C4-09 -- the unusual four-knot system
of Steidel \etal\ (1996b).  We confirm the result of Zepf \etal\
(1996) that the \lya\ emission line is clearly resolved into two
components separated by 14 \AA, or 820 $\kms$ in the rest
frame\footnote[6]{Note that we find redshifts $z=3.210, 3.222$,
somewhat lower than the redshift $z=3.226$ reported by Steidel
\etal\ but in good agreement with Zepf \etal} (Fig. 2).  We can use
this separation to estimate the mass of the system under the
assumption that the split is due to gravitationally-induced motions.
The slitlet was aligned at $PA=137\deg$, constrained by the two
candidate galaxies on either side of C4-09.  The longest projected
separation of the four knots of emission along the slit, then, is
0\decsec9 or $\sim 11\ h_{50}^{-1}$ kpc ($q_0 = 0.05$) (although the
seeing disk smeared the image sufficiently that our spectrum has no
spatial information).  For a circular velocity of 100 $\kms$ at a
radius $r=5.5$ kpc, we then infer a gravitational mass of $\sim
2\exp{11}\ \msun$, comparable to local massive galaxies.

The moderate star formation rates implied by the UV fluxes may be
difficult to reconcile with the rapid, early formation of massive
spheroids usually inferred from stellar population studies of local
galaxies (\eg, Bower, Lucey, \& Ellis 1992).  At a constant star
formation rate of $10\ \sfr$, the high-redshift HDF galaxies would
require $10^{10}$ yr to produce a $10^{11}\ \msun$ spheroid.
Conversely, the mass-to-light ratios for active starbursts are as low
as $M/L = 0.1$ (Guzm\'an \etal\ 1996), implying typical masses $M
\sim\ 10^{10}\ \msun$ for our calculated luminosities.

More clues to the nature of the $z=3$ galaxies come from their space
density compared to local galaxies today.  Considering only the 11
confirmed high-redshift galaxies presented here plus the 5 observed
by Steidel \etal\ (1996b) yields at least 16 objects in the range
$2.2 < z < 3.5$ in the HDF.  The 4.7 arcmin$^2$ field constitutes a
volume of 68000 $\ h_{50}^{-3}$ comoving Mpc$^3$ for $q_0=0.05$ or
14000 $\ h_{50}^{-3}\ \mpc^3$ for $q_0=0.5$ over that redshift
range, so the space density of high-redshift star-forming galaxies
is $n \geq 2.4\exp{-4} \ h_{50}^3\ \mpc^{-3}$ for $q_0=0.05$ or $n
\geq 1.1\exp{-3}\ h_{50}^3\ \mpc^{-3}$ for $q_0=0.5$.  The latter
value is close to twice the density of $L \geq{\rm L}^*$ galaxies in
the local Universe, $n(L\geq \rm L^*) = 6 \exp{-4}\ h_{50}^{3}
\mpc^{-3}$ (where we have integrated the bright end of the luminosity
function derived by Lin \etal\ 1996 from the Las Campanas Redshift
Survey), and 3 - 4 times higher than the values found by Steidel
\etal\ (1996a) from ground-based $UGR$ photometric selection and Keck
spectroscopy of a slightly brighter sample.  This higher density
reflects the additional 1 magnitude of depth in our sample; the
additional confirmed $U$-dropouts; and the $B$-dropouts, which alone
provided 27\% of our confirmed sources, not including those objects
that were also $U$-dropouts.

If the 16 objects we did {\em not} observe (due to time constraints),
which have a color and magnitude distribution similar to the observed
sample, also have a 44\% success rate of redshifts $2.2 < z < 3.6$,
the additional seven high-redshift objects would boost the number
density in the HDF by almost 50\%.  Furthermore, since none of the
non-confirmed objects we observed is inconsistent with being in that
redshift range, the total list of 46 high-redshift candidate galaxies
could yield a number density as much as three times higher than the
figure quoted above.  If indeed all of these sources are the
precursors of today's $L >$ L$^*$ ellipticals and spheroids of
spirals, then substantial merging and/or fading must have taken place
between then and now to reduce their numbers and reconcile them with
the local luminosity function.

Lanzetta \etal\ (1996) estimate the surface densities of galaxies in
the HDF as a function of redshift and apparent magnitude based on
their photometric properties.  The surface density of the DEEP sample
of confirmed high-redshift galaxies, coupled with Steidel \etal's
(1996b) and normalized to 1 arcmin$^2$, agrees reasonably well with
those estimates except in the lowest of the three redshift bins we
targeted.  For $I_{814,AB}<26$, the predictions are $8.1\pm0.7,
1.9\pm0.3,$ and $2.1\pm0.3$ galaxies arcmin$^{-2}$ in the redshift
ranges 2.0 - 2.5, 2.5 - 3.0, and 3.0 - 3.5, respectively.  We find
0.6, 1.5, and 1.3 confirmed galaxies arcmin$^{-2}$ in those ranges,
consistently below the $1 \sigma$ error boundary, but since we expect
many or most of the unconfirmed galaxies also to lie in those ranges,
the actual surface densities probably match or even exceed the
estimates of Lanzetta \etal\ in the higher redshift bins.

The integrated star formation rate for the 11 confirmed high-redshift
galaxies is $SFR = 200\ h_{50}^{-2}\ \sfr$ ($q_0=0.05$) or $72\
h_{50}^{-2}\ \sfr$ ($q_0=0.5$) in the volumes quoted above.  This
amounts to $2.9 - 5.1 \exp{-3} h_{50}^{-1}\ \sfr {\rm Mpc}^{-3}$ (for
$q_0=0.05 - 0.5$), about 10 - 15\% of the local value of $0.04\ \sfr
{\rm Mpc}^{-3}$ measured by Gallego \etal\ (1996; note that we have
divided the integrated SFR from Gallego \etal\ by 3.3 to transform
from a Scalo IMF to a Salpeter IMF, \eg, Kennicutt 1983).  If we
include the objects from Steidel \etal\ (1996b), the measured
integrated SFR at $z=3$ approaches one-half the local value for
$q_0=0.5$; including the extinction correction mentioned in \S 5.1
would bring the two values into rough equilibrium.  The $z=3$ value is
still an order of magnitude lower than the star formation density
measured at $z=1$ by Lilly \etal\ (1996) from the CFRS, but of course
the $z=3$ estimate is a lower limit based only on the bright end of
the luminosity function of star-forming galaxies at $z=3$; as Gallego
\etal\ show for the local Universe, as much as 90\% of the integrated
SFR can come from the faint end ($L < \rm L^*_{\alpha}$) of the
\ha\ luminosity function.

There is a hint from the 16 confirmed high-redshift objects in the HDF
(the present sample plus that of Steidel \etal\ 1996b) of luminosity
or density evolution with redshift.  If we divide the combined samples
into three redshift ranges 2.20 - 2.68, 2.68 - 3.11, and 3.11 - 3.50,
each with volume $V \sim 22,000\ h_{50}^{-3}\rm{Mpc}^3$ ($q_0=0.05$),
then we find 4, 6, and 6 objects, respectively, in each bin.  However,
if we restrict the sample to the 12 objects brighter than $M_B=-22.3$,
then we find only one object in the low-redshift bin, but six and five
objects in the medium- and high-redshift bins, respectively.  This
suggests that luminous star-forming galaxies have either faded or
merged or both during the $\sim 2$ Gyr (for $q_0=0.05, H_0=50\
\kmsmpc$) between $z=3.5$ and $z=2.2$.  Alternatively, they may
progressively extinguish their UV light as massive stellar processing
causes an increase in metallicity and dust.  Clearly this will be a
fascinating line of inquiry to pursue with larger datasets that sample
more densely the luminosity function at high redshift.

\subsection{A Low-Mass Starburst Scenario}

To help guide the discussion and interpretation of the star-forming
galaxies at $z\sim3$ in the HDF, including those observed by Steidel
\etal\ (1996b), we summarize their salient observed characteristics
and then speculate on the ultimate fate of these distant galaxies and
their possible connections to normal galaxies in the local Universe:

\begin{itemize}

\item Typical $B$-band luminosities are 1 - 2 magnitudes brighter than
present day L$^*$.

\item Median half-light radius is small: $r_{1/2}=3.6 h_{50}^{-1}$ kpc for
aggregate systems and $r_{1/2}=2.7 h_{50}^{-1}$ kpc for subclumps
($q_0=0.05$; smaller by 0.6 times for $q_0=0.5$).

\item Morphologies are varied, with multiple knots of emission
and diffuse wispy tails that imply non-relaxed systems.

\item Star formation rates are modest: 7 - 24 $h_{50}^{-2}\sfr$ for
$q_0=0.05$ (3 - 9  $h_{50}^{-2}\sfr$ for $q_0=0.05$; not corrected for
dust extinction).

\item \lya\ emission is weak when it is present at all, and the
observed profile is narrow, $FWHM_{\rm med} = 7$ \AA, implying
$\sigma_{\rm med} = 140\ \kms$.

\item Stellar and interstellar absorption lines of heavy elements are
similar to those seen in the spectra of nearby starburst and WR
galaxies, but narrow, weak stellar-wind profiles and equivalent widths
may imply low metallicity.

\item The comoving volume density, including the 5 objects reported by
Steidel \etal\ (1996b), is $n \geq 2.5\exp{-4} h_{50}^3 {\rm
Mpc}^{-3}$ for $q_0=0.05$, or $n \geq 1.2\exp{-3} \ h_{50}^3\
\mpc^{-3}$ for $q_0=0.5$, comparable to the local volume density of
galaxies with $L > {\rm L}^*$.  The volume density could be 3 - 4
times higher if most of the remaining candidates are also at $2.2 < z
< 3.5$. 

\item Strong clustering is not observed.

\item  Star-forming galaxies are seen from $z=2.2$ to $z=3.5$,  the
range to which our study was sensitive; the redshift distribution
appears to be slightly weighted towards higher redshifts.

\end{itemize}

We consider three possible scenarios of galaxy formation and evolution
that might explain the present observations.  In the first scenario,
the luminous, compact, star-forming objects we observe represent the
{\em cores} of {\em massive} spheroids of today's $L> {\rm L}^*$
elliptical and spiral galaxies.  In this picture, which is perhaps
similar to what Steidel \etal\ (1996a,b) have proposed, the sites of
star formation lie near the centers of deep potential wells that are
continuously funnelling in new gas and processing it into stars at
roughly $10\ \sfr$ for many Gyr, eventually building a stellar mass
$M\sim 10^{10}\ \msun$.  In support of such a view is the apparent
agreement between the half-light radii of high-redshift galaxies and
those of local spheroids and bulges, as noted above, and the similar
comoving number densities of high-redshift galaxies and $L>{\rm L}^*$
galaxies today.

It seems clear from the HDF redshift distribution that the
high-redshift population is a {\it field} population, not the
precursor of clusters.  The random statistics of probing high
redshifts favor field galaxies overwhelmingly in any case.  That,
together with the large number density, suggests that if they are
indeed forming spheroids, then these objects are not the spheroids of
massive ellipticals but rather the spheroids/bulges of early-type
spirals. This is consistent with the half-light radii, which match
those of local bulges but are smaller than those of giant ellipticals
(\eg, Bender et al. 1992).

However, the size agreement may be somewhat superficial: If these
early phases of galaxy evolution show low-metallicity stars, analogous
to Pop II stars in the Milky Way, then we should properly compare with
the radii of local Pop II components. This is larger than $r_{1/2}$
given above -- in the Milky Way, for example, the Pop II component of
globular clusters with $Z < 0.1$ has a median radius of 10 kpc (Harris
and Racine 1979) and extends to $\sim 50$ kpc, very much larger than
the median half-light radius $r_{1/2}$ = 3.6 kpc we observe for even
the {\em aggregate} systems.  The number density agreement may be
somewhat superficial as well, since significant merging can easily cut
the numbers by factors of several and/or induce additional luminous
bursts of star formation, and passive fading of the stellar components
can dim the objects by 5 magnitudes (see previous section).

A second possible scenario borrows from models of hierarchical
clustering, which predict that highly overdense systems like those
believed to form spheroids and bulges (Blumenthal et al. 1984) begin
to collapse with many subclumps (Kauffmann, White, \& Guiderdoni 1993;
Navarro, Frenk, \& White 1995; Cole \etal\ 1994).  A subclump model
has also been advanced by Searle and Zinn (1978) for the formation of
the metal-poor spheroid of the Milky Way.  We might therefore expect
to see a {\it population} of subclumps in every high-redshift
proto-spheroid, distributed over a volume $\sgreat\ 10$ kpc in radius
, or 1 - 2\sec\ on the sky.  This is not in general seen (although
there are a handful of cases in which companion blobs at radii $\sim 1
- 2$\sec\ have colors that suggest similar redshift).  This seeming
contradiction might be accounted for if the bright phase of each
individual blob lasted for only a few $\times 10^8$ yr, so that, on
average, only one blob is seen at any one time.  We then have what
might be termed a ``Christmas tree'' model, wherein small individual
star-forming blobs come and go within a much larger -- and largely
invisible -- spheroid structure.  It will be important to gather
dynamical evidence to test this picture.

Finally, we suggest a third alternative in which the $z=3$ galaxies
are relatively low-mass {\em isolated} knots of star formation similar
to current-day H~II galaxies, 10 - 100 times less massive than typical
L$^*$ disk spheroids or ellipticals.  In this scenario, rather than
forming stars at a steady rate over many Gyr, the galaxies at $z=3$
would be converting gas to stars in small components through more
intense bursts with low mass-to-light ratios, such as the CNELGs
observed at lower redshifts with $M/L\sim0.1$ (Guzm\'an \etal\ 1996).
This would naturally explain the apparently high number density of
galaxies at $z=3$ in the HDF, since relatively low-mass objects would
be easily included in our sample due to their intense brightness.  The
diffuse, wispy tails and diverse morphologies of many of our sample
sources indicate that a large supply of gas is generally available for
future star formation nearby; as star formation progresses in the
luminous, compact knots, gas may be used up locally or expelled
through supernova-driven winds (e.g.  Guzm\'an \etal\ 1996; Lehnert \&
Heckman 1996), star formation may taper off exponentially, and new
regions of star-forming activity may spring up in other dense pockets
of gas.  The end products of such a process could be low-mass
spheroidal galaxies such as those observed in the Local Group.  The
diverse morphologies, moderate star formation rates, and small sizes
all seem to favor the ``Christmas tree'' picture or this low-mass
burst scenario.

Of course, the actual picture may be some combination of those three
descriptions: extensive merging of clumps could produce a wide range
of galaxy types and masses by the present epoch.  Such a scenario is
supported by recent N-body and semi-analytical models of galaxy
formation and evolution (\eg, Kauffmann \& White 1993), which favor
complex and rich merging histories for most massive galaxies.  The
small individual knots that we observe within regions only a few kpc
across in four confirmed high-redshift systems are also likely to
merge soon: even assuming a mass-to-light ratio as low as $M/L=0.1$,
appropriate for an intensely star-forming system (Guzm\'an
\etal\ 1996), the dynamical time for an L$^*$ object 3 kpc across
is $\sim 0.1$ Gyr.  

One important caveat to keep in mind when interpreting rest-UV
morphologies, of course, is that those morphologies may bear little
resemblance to the rest-optical view of galaxies with which we are
familiar in the local Universe.  Even ``grand design'' spiral galaxies
often display barely recognizable, scattered clumps of star-forming
regions when observed at 1200 \AA\ (O'Connell \& Marcum 1996).  We are
certainly observing only the most intense bursts of star formation at
the location and epoch surveyed by the HDF; there are, no doubt,
complex underlying mass structures and fainter star formation sites
lurking below our current detection limits.

We cannot measure the masses of $z=3$ galaxies directly given the
present data.  Therefore, while we find no direct evidence that the
$z=3$ galaxies discussed here are massive ($M > 10^{10} \msun$), there
is also no evidence that they will not eventually merge with similar
objects to {\em become} massive galaxies today.  In any case, whether
any of these particular explanations is correct, it is clear that the
morphologies, sizes, spectra, and numbers of the high-redshift
galaxies contain key information on the structure of proto-galaxies.

\begin{acknowledgements}

We are indebted to the staff of the W. M. Keck Observatory for their
expert assistance in obtaining the data presented here.  Thanks are
due to C. Leitherer for providing the digital spectrum of NGC 1741,
and to C. Steidel for helpful comments and for pointing out an error
in one of the redshifts.  We appreciate a careful reading by K. Wu.
Support for this work was provided by the National Aeronautics and
Space Administration through grant numbers AR06337.08-94A and
AR06337.21-94A and Hubble Fellowship grant HF-1048.01-93A from the
Space Telescope Science Institute, which is operated by the
Association of Universities for Research in Astronomy, Inc., under
NASA contract NAS5-26555, and by the National Science Foundation
through grant numbers AST-91-20005 and AST-95-29098.  JG acknowledges
the partial financial suport from Spanish MEC grants PB89-124 and
PB93-456 and a UCM del Amo foundation fellowship.  CG acknowledges
support from an NSF Graduate Fellowship.  This research has made use
of the NASA/IPAC Extragalactic Database (NED), which is operated by
the Jet Propulsion Laboratory, Caltech, under contract with the NASA.
 
\end{acknowledgements}

\vfill\eject

\appendix 

\section{Notes on individual objects}

\centerline{Listed in order of increasing Right Ascension, as in Table 1.}

\subsection {\bf hd4\underbar{~}0259\underbar{~}1947}  A clearly disk-shaped
galaxy 2\sec\ from a larger and brighter disk.  The object is barely
visible in the F300W image, but is a solid $U-$dropout, implying
$z>1.7$ (for a \lya\ break) or, more likely given the large $U-B$
color, $z\ \sgreat\ 2.6$ (for a Lyman continuum break).  If confirmed
to be at such a high redshift, this would certainly be one of the
earliest-formed disk systems known.  The larger disk nearby is
considerably redder and therefore probably at lower redshift.

\subsection {\bf hd4\underbar{~}1076\underbar{~}1847}  The brightest galaxy
in the sample, but uncertain redshift.  Mg~II absorption at $z=0.879$
and $z=1.010$ is detected in the spectrum (cf. \S 4), presumably due
to the galaxies at those redshifts 1-2\sec\ to the west and south,
respectively.  The spectrum shows no definite
\lya\ break or emission lines above 4580 \AA, so the redshift is most
likely $1.01 < z < 2.8$.  There is a possible weak emission line at
5433 \AA.  The color matches that of confirmed galaxies at $z\sim2.5$;
the absorbing galaxies at $z=0.879$ and $z=1.010$ have very different
colors.  After removing the Mg~II absorption lines from the spectrum,
we find a cross-correlation peak at $z=2.41$, but no other features to
confirm that redshift.  A weak absorption lines at 4884 \AA\ could
match \civ\ at $z=2.15$ (or Al~II at $z=1.93$), however, and we adopt
that as a tentative redshift.

\subsection {\bf hd4\underbar{~}1229\underbar{~}1791 ($z=0.483$)}  A
diffuse, low-surface brightness galaxy; the only object in the sample
with a confirmed redshift $z<2$.  The source is only 4\sec\ and 7\sec\
from galaxy C4-09 at $z=3.22$ of Steidel \etal\ (1996b) and
hd4\underbar{~}1076\underbar{~}1847 at $z\geq1$, respectively.
Evidently the 4000 \AA\ break, redshifted to $\sim 6000$ \AA, allowed
the object to be included in the $B-$dropout sample.  The single
emission line in the blue spectrum (shown in Fig. 2) is apparently
\oii, with a rest equivalent width $W_{\rm [O II]} = 39$\AA.  The red
spectrum shows possible \hb\ and faint \oiii\ emission lines, and the
strong $U$-band detection ($U_{300,AB}=26.86$) further demonstrates
that the redshift must be $z<2.6$.

\subsection{  hd2\underbar{~}2030\underbar{~}0287 ($z=2.267$):}  We
identify the strong emission line as \lya; weak absorption lines match
the expected wavelengths of Si~II, C~II and C~IV.  If the emission
line were \oii\ at $z=0.066$, the absolute magnitude would be
$M_B\sim-11.7$ and the [O~II] rest equivalent width would be $W_0>100$
\AA, high compared to local samples (\eg, Terlevich \etal\ 1991;
Broadhurst, Ellis, \& Shanks 1992) although not implausible.

\subsection{ hd4\underbar{~}0818\underbar{~}1037}  The spectrum  shows
reasonably strong continuum with a strong absorption line at 5066 \AA,
identified by Cohen \etal\ (1996) as \civ\ at $z=2.268$.  However, no
other features matching that redshift are evident, and we list in
Table 2 a tentative alternate redshift $z=2.035$ based on the best
peak of the cross-correlation {\em with the combined spectrum shown in
Fig. 4}; this corresponds to the absorption line being Al~II.

\subsection{ hd2\underbar{~}1881\underbar{~}0374}  This isolated, compact
source shows a healthy $U-B$ break but is detected in the F300W image,
so the redshift is most likely $z<2.6$.  The spectrum shows some broad
features but we were unable to identify any with certainty.

\subsection{ hd4\underbar{~}1994\underbar{~}1406}  Although the
cross-correlation peak for this compact galaxy is at $z=3.630$ --
higher than any other object in the sample, confirmed or unconfirmed
-- we regard this redshift as suspect since the object is detected in
the F300W image.

\subsection{  hd2\underbar{~}1410\underbar{~}0259  ($z=3.160$)}  A
complex source with a bright asymmetrical core and extended plumes of
emission, all with roughly the same color.  We identify the emission
line as \lya, and there is good agreement with weak absorption lines
matching Si~II, O~I, and C~II.  If the emission line were instead
\oii\ at $z=0.359$, then we would expect to see \hb\ and \oiii\ in the
blue spectrum and/or \ha\ in the red spectrum; no such lines are seen
(although the red spectrum is heavily contaminated by night sky
emission lines).

\subsection{ hd4\underbar{~}1486\underbar{~}0880}  One of the most
symmetrical profiles in the sample.  The best peak of the
cross-correlation with the spectrum of NGC~1741 is at $z=2.952$, but
we list instead $z=2.47$ based on the general shape and possible
agreement with a few stellar and interstellar absorption lines.

\subsection{ hd2\underbar{~}0853\underbar{~}0319}  A very faint, diffuse
galaxy, with almost undetectable continuum and no emission lines or
strong breaks in the blue or red spectrum.  The cross-correlation
function peaks at $z=3.35$, but the object is weakly detected in the
F300W filter, so it may in fact lie at $z<2.6$; the morphology and
color are indeed similar to those of
hd4\underbar{~}1229\underbar{~}1791 ($z=0.483$), the only galaxy with
a confirmed redshift $z<2$.

\subsection{  hd4\underbar{~}0367\underbar{~}0266  ($z=2.931$)}  Two
nearly equal knots of emission are roughly aligned with a clumpy,
diffuse tail, all of approximately the same color.  The broad \lya\
absorption trough, strong \lya\ break, and strong, sharp absorption
lines of many interstellar and stellar ions provide a secure redshift.
The disk galaxy a few arecsec to the southwest has a different color
and therefore probably a significantly different redshift.

\subsection{   hd4\underbar{~}2030\underbar{~}0851  ($z=2.980$)}  The
least secure of the confirmed subsample.  The redshift is based on the
break at 4850\AA, which we identify as \lya, and is supported by weak
absorption line matches with the wavelengths expected for
Si~II~1526\AA\ and C~IV~1550\AA\ and by the object's non-detection in
the F300W filter.  The object 1\sec\ to the northeast is almost
certainly at lower redshift, judging from its red color.

\subsection{ hd4\underbar{~}1341\underbar{~}0299} A compact object with a
close companion of considerably different color, and therefore
probably a chance projection rather than physical association.  The
spectrum shows one possible weak emission line at 5334\AA, which could
match \lya\ at $z=3.379$, but no \lya\ break or interstellar
absorption lines are present; we list in Table 2 $z=2.31$ based on a
peak in the cross-correlation function (a weaker peak appears at
$z=3.35$).

\subsection{ hd2\underbar{~}1739\underbar{~}1258}  Morphologically the
most complex object in the sample.  Numerous bright knots and
filaments with a broad range in color are arranged in a rough line
3\sec\ long, and a redder clump of emission -- possibly at different
redshift -- lies only 1\sec\ to the west.  Our spectrum was obtained
with the slitlet aligned with the object's major axis.  Despite the
object's relative brightness, the blue spectrum failed to reveal any
identifiable features.  The red spectrum shows a possible faint
emission line at 6850 \AA\ that would match \oii\ at $z=0.838$, but no
other features support that redshift, and the line is not convincing.
We adopt $z=2.72$, the cross-correlation peak that provides the best
matches with \lya\ and \civ\ absorption.

\subsection{  hd3\underbar{~}0408\underbar{~}0684  ($z=3.233$)}  We
identify the strong emission line as \lya.  No absorption lines are
detected, but there is a clear break across the emission line and a
strong $U-$band limit that support the identification.  

\subsection{ hd2\underbar{~}1410\underbar{~}1282}  The faint continuum in
our spectrum yielded no secure features; the redshift listed in Table
2 is from the cross-correlation spectrum, supported by the large $U-B$
color.

\subsection{  hd2\underbar{~}0698\underbar{~}1297  ($z=3.430$)}  The
highest reliable redshift in the sample is based on the \lya\ emission
line and continuum break, and good matches of absorption lines with
Si~II and O~I.  The \lya\ line appears slightly redshifted ($z=3.439$,
perhaps due to internal self-absorption) with respect to the
absorption lines, which we use to derive the exact redshift.  The
images show two close knots of emission adjacent to a wispy cloud, all
with similar colors.  The source is only 4\sec\ (48 $h_{50}$ kpc,
$q_0=0.05$) projected separation from
hd2\underbar{~}0705\underbar{~}1366 ($z=3.368$), but the redshift
difference corresponds to more than 20,000 $\kms$ in the rest frame,
so the likelihood of physical association is low.  However, there are
two possible absorption lines in the spectrum that show rough
agreement with Si~II $\lambda1260$ and O~I $\lambda1303$ at $z=3.368$,
presumably due to hd2\underbar{~}0705\underbar{~}1366.

\subsection{hd2\underbar{~}0705\underbar{~}1366  ($z=3.368$)}  The
strong emission line and break are almost certainly \lya.  A possible
weak absorption line at 4482 \AA\ matches the wavelength expected for
Ly$\beta$ at the \lya\ emission redshift, but the continuum is
extremely weak here and the absorption line is probably spurious.
Despite some faint extended emission, the bright core gives this
galaxy the smallest half-light radius in our sample,
$r_{1/2}=$0\decsec14, indistinguishable from a point source.

\subsection{ hd2\underbar{~}1918\underbar{~}1912}  A Galactic star, also
observed by Steidel \etal\ (1996); it was included in our sample
through an oversight of that fact.

\subsection{ hd2\underbar{~}0434\underbar{~}1377  ($z=2.991$)}  This
compact, slightly elongated source shows the strongest absorption
lines in the sample.  The \lya, Si~II, O~I, and C~II absorption
lines make the spectrum virtually identical to that of NGC~1741, the
local comparison starburst galaxy used in the cross-correlations.

\subsection{   hd2\underbar{~}1359\underbar{~}1816  ($z=3.181$)}  The
redshift is based on the excellent agreement of the clear break at
5100\AA\ with \lya\ and of several strong absorption lines with common
UV stellar and interstellar features.

\subsection{   hd2\underbar{~}0624\underbar{~}1688  ($z=2.419$)}  Three
very close knots of emission are contained within 1\sec.  The redshift
is based on the \lya\ line; no absorption lines are detected.  Since
no other emission lines are visible in either the blue or red
spectrum, the only other plausible assignment for the emission line --
\oii\ -- is highly unlikely.

\subsection{ hd3\underbar{~}1455\underbar{~}0430}  A tight, symmetrical
profile with a faint companion less than 1\sec\ to the north, and
several objects with comparable brightness but dissimilar colors (and
therefore redshifts) within 5\sec.  Continuum is quite strong in our
spectrum, and there are two absorption lines at 5097 \AA\ and 5648
\AA, but we were unable to make a positive identification.  We adopt
$z=2.644$ based on the general shape and the location of weak features
matching
\civ\ and interstellar UV lines.

\subsection{ hd2\underbar{~}0725\underbar{~}1818 ($z=2.233$)}  This
asymmetrical galaxy lies only 1\sec\ to the southwest of
hd2\underbar{~}0743\underbar{~}1844, and the two sources' colors are
almost identical.  The excellent agreement of strong absorption lines
with the wavelengths of Si~II, O~I, C~II, and Al~II and the possible
emission line of C~III] make the redshift assignment secure.

\subsection{ hd2\underbar{~}0743\underbar{~}1844}  Despite the apparent
similarities and proximity to hd2\underbar{~}0725\underbar{~}1818,
this galaxy's spectrum yields no secure features.  We adopt the
redshift $z=2.390$ given by the best peak of the cross-correlation
function.

\newpage

\begin{figure}
\vspace*{8in}
\begin{minipage}{7in}
\includegraphics{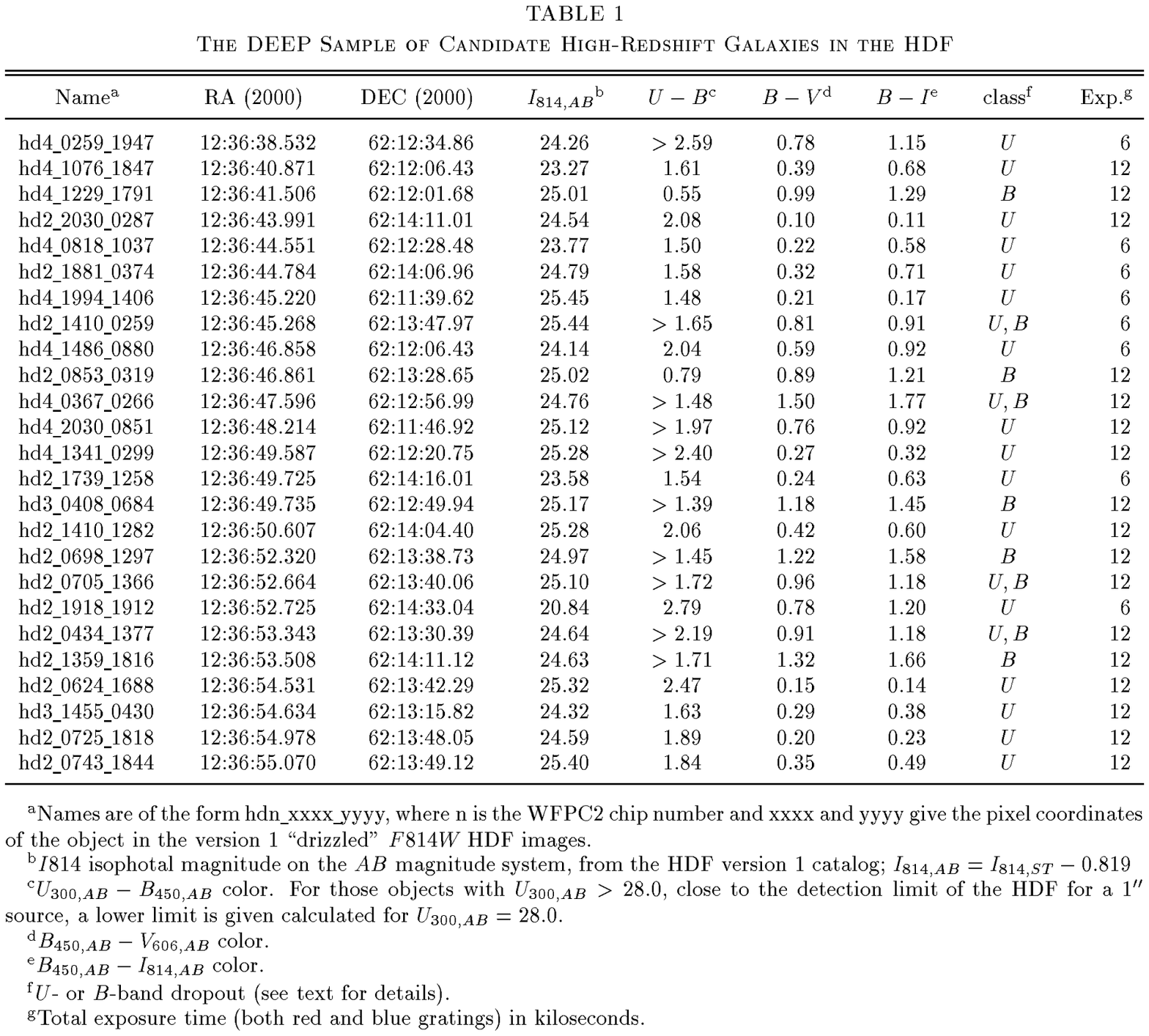}
\end{minipage}
\end{figure}

\newpage

\begin{figure}
\vspace*{8in}
\begin{minipage}{7in}
\includegraphics{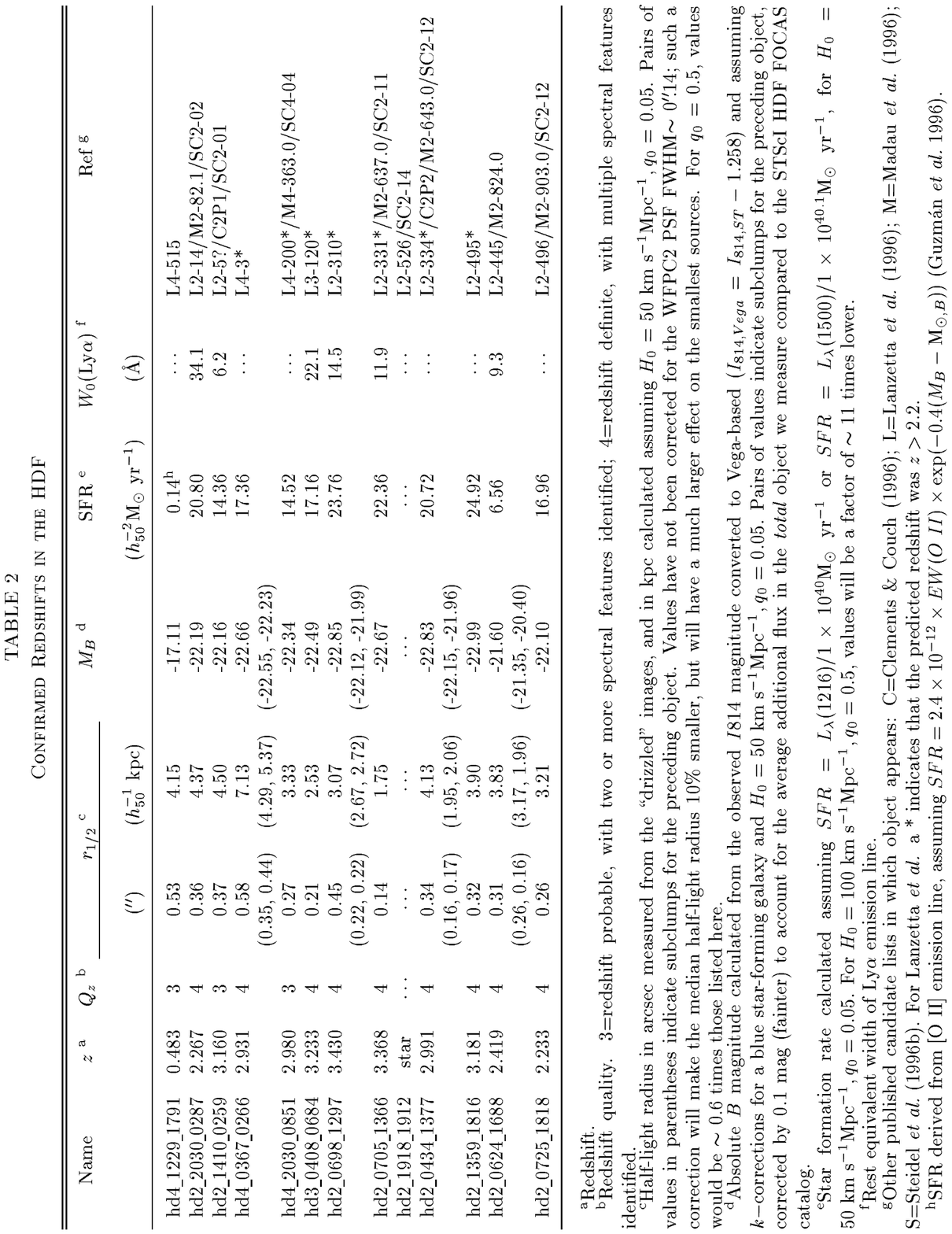}
\end{minipage}
\end{figure}

\newpage

\begin{figure}
\vspace*{8in}
\begin{minipage}{7in}
\includegraphics{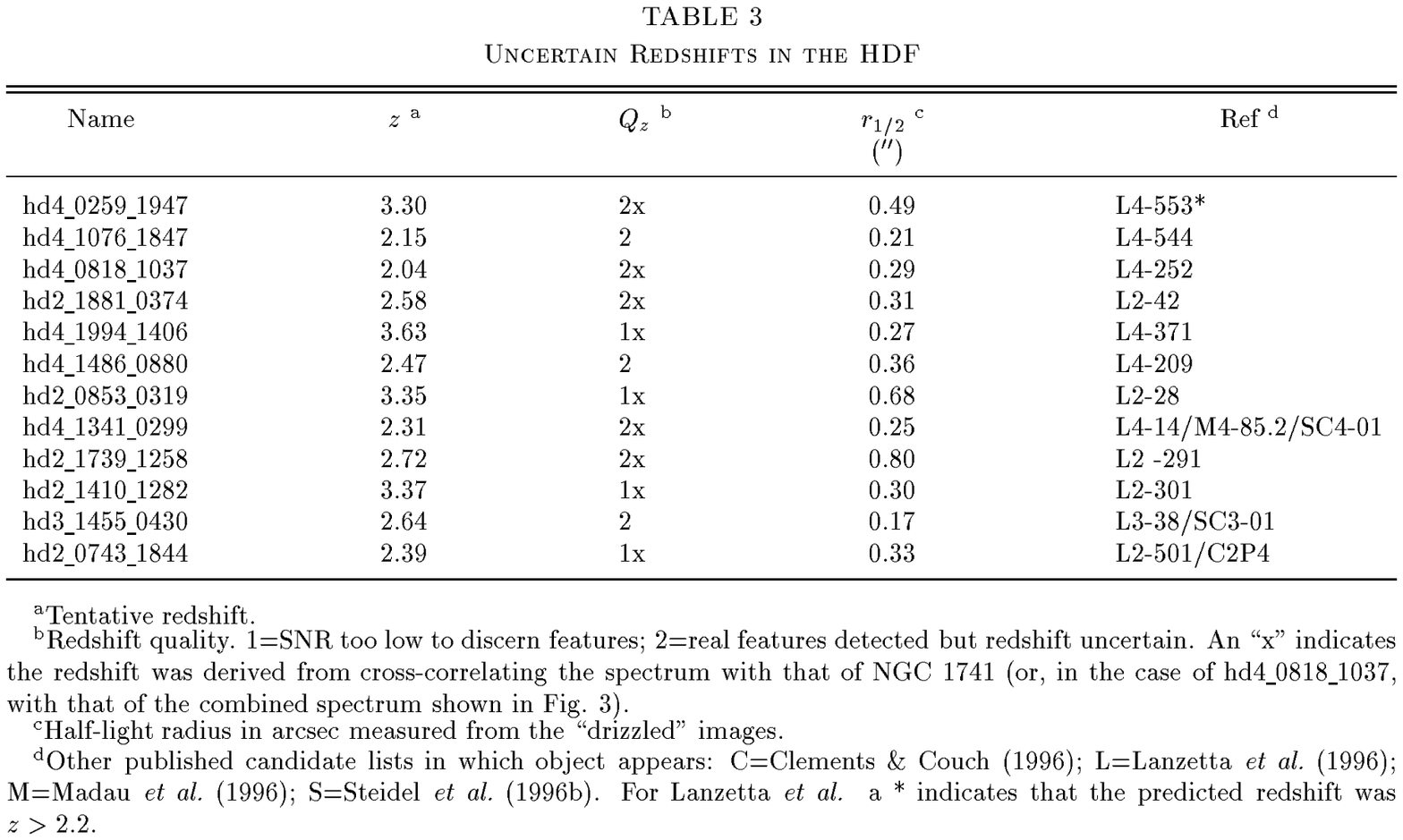}
\end{minipage}
\end{figure}

\newpage

\centerline{\bf Figure Captions}

\noindent {\bf Figure 1.}  {The DEEP sample of candidate high-redshift
galaxies in the Hubble Deep Field.  Only the 24 galaxies observed at
Keck are shown.  The images were excised from the version 1
``drizzled'' HDF F450W, F606W, and F814W images, combined using the
blue, green, and red color guns, respectively.  The top twelve objects
have confirmed redshifts, labeled in the upper right corner, and are
presented in order of increasing redshift.  The effect of redshift on
the objects' apparent colors for $2<z<4$ is easily seen in this
representation, with the lowest redshift galaxies appearing blue and
the highest, green.  The bottom twelve objects do not have confirmed
redshifts, and are presented in order of increasing $B-V$ color.
North is up, east is to the left, and each box is 10\sec\ $\times$
10\sec.}

\noindent {\bf Figure 2.} {{\bf (a)} One-dimensional
optimally-extracted spectra of the twelve galaxies with confirmed
redshifts, arranged in order of increasing redshift.  Only the first
galaxy has $z < 2.2$.  Each spectrum has been smoothed by 11 pixels to
reduce the noise.  On the bottom of each panel of four spectra is the
GHRS spectrum of the nearby irregular starburst galaxy NGC 1741 from
Conti \etal\ (1996), redshifted for comparison with the HDF galaxies.
The wavelengths of several rest-UV spectral features commonly seen in
actively star-forming regions are indicated.  In the spectrum of
hd2\underbar{~}0698\underbar{~}1297, we have labeled the wavelengths
of possible redshifted Si~II $\lambda1260$ and O~I $\lambda1303$
absorption at $z=3.368$ from galaxy
hd2\underbar{~}0705\underbar{~}1366, 4\sec\ away.  {\bf (b)}
Two-dimensional spectra showing the region near \lya\ for the six
objects showing \lya\ emission.  Note the significant drop in
continuum across the \lya\ lines.  Also shown are the \oii\ emission
line at $z=0.483$ from hd\underbar{~}1229\underbar{~}1791 and the
double \lya\ emission line from object C4-09 of Steidel \etal\
(1996b).  The small horizontal lines labeled a, b, and c indicate the
spectra of hd2\underbar{~}0725\underbar{~}1818 ($z=2.233$),
hd2\underbar{~}0743\underbar{~}1844, and
hd4\underbar{~}1076\underbar{~}1847 ($z \geq 1.01$), respectively.
Each image section is 500 \AA\ on the horizontal axis, with
wavelengths increasing to the right; tick marks on the vertical axis
are at 1\sec\ and 5\sec\ intervals.}

\noindent {\bf Figure 3.} {Spectrum of
hd4\underbar{~}1076\underbar{~}1847, the brightest galaxy in the
sample.  Despite the strong continuum and the presence of several
absorption lines and a possible emission line (most of them more
clearly visible in the two-dimensional spectrum), we are unable to
determine a secure emission redshift.  Two sets of Mg~II absorption
lines at $z=0.879$ and 1.010 from galaxies $<$ 2\sec\ from the line of
sight are indicated, and the absorption line tentatively identified as
C~IV at $z=2.15$ (or Al~II at $z=1.93$) is labeled.  Other absorption
lines may be due to additional intervening systems at different
redshifts.

\noindent {\bf Figure 4.} { Average spectrum of the 11 confirmed high-redshift
galaxies in the HDF from the DEEP sample plus C4-09 of Steidel \etal\
(1996b).  Note the weak \lya\ emission line, the strong continuum
break and broad absorption trough at \lya, the weak emission lines of
He~II and C~III], and the absorption lines of Ly$\gamma$, Ly$\beta$,
Si~II, O~I, C~II, Si~IV, C~IV, Fe~II, and Al~III.  The high-ionization
lines Si~IV and C~IV are weak and narrow compared to those seen in the
spectra of high-metallicity hot stars (Walborn
\etal\ 1996), implying that the average galaxy at $z\sim3$ is not
heavily metal-enriched.}

\noindent {\bf Figure 5.} {Redshift vs. $I_{814,AB}$ for the confirmed
sample.  Also shown for comparison are the faint magnitude limit of
the Canada France Redshift Survey (Lilly \etal\ 1996); three lines
corresponding to unevolved L*, 10 L*, and 100 L* blue star-forming
galaxies; the five confirmed high-redshift objects in the HDF observed
by Steidel \etal\ (1996b; $\times$ symbol); two galaxies that have
been interpreted as possible primeval galaxies, IRAS 10214+4724
(Rowan-Robinson \etal\ 1991) and CNOC cB58 (Yee \etal\ 1996), and the
gravitationally lensed high-redshift galaxies discovered by Trager
\etal\ (1996; open circles), although all four of those galaxies may be
gravitationally lensed and thus magnified by up to a factor of 10; a
\lya- and \ha-emitting galaxy at $z=2.5$ found near a QSO
absorption-line cloud (Malkan \etal\ 1995; open box symbol); and two
objects at $2<z<3$ in the Flanking Fields that were selected based on
their compact morphology and high surface brightness
(``serendipitous''; see Phillips \etal\ 1996 for more details).  The
present sample probes significantly fainter than most previous studies
of high-redshift galaxies: the confirmed sample comprises objects that
are barely more luminous than L* galaxies today.}

\noindent {\bf Figure 6.} {Color-color distributions of galaxies in the
HDF.  {\bf (a)} $B_{450,AB}-I_{814,AB}$ vs.
$U_{300,AB}-B_{450,AB}$. {\bf (b)} $V_{606,AB}-I_{814,AB}$
vs. $B_{450,AB}-V_{606,AB}$. The 11 confirmed high-redshift galaxies
are indicated by open hexagons, the 12 galaxies with unconfirmed
redshifts are indicated by open circles, the Galactic star is
indicated by a star symbol, the galaxy at $z=0.483$ is indicated by an
open triangle, and additional galaxies from the HDF WF2 catalog that
we did not observe are shown by open squares, to indicate the field
galaxy population.  (Note that we have omitted field galaxies from the
WF3 and WF4 chips for clarity; the actual density of field galaxies
compared to the observed sample is thus about three times higher than
shown.)  Candidate galaxies with $U_{300,AB}\geq28.0$ are shown at a
conservative lower limit corresponding to $U_{300,AB}=28.0$ and are
indicated by a vertical line.  Also shown ($\times$ symbol) are the
five confirmed high-redshift galaxies of Steidel \etal\ (1996b).  The
solid lines indicate the regions of color-color space we used to
select our $U$-band ({\em a}) and $B$-band ({\em b}) dropout
candidates.  The regions chosen by Madau \etal\ (1996) to select
high-redshift candidates are indicated by dotted lines.  There are
five objects that do not satisfy those criteria, even after including
the higher lower limits of Madau
\etal, and yet are confirmed to have $z>2.9$.}

\noindent {\bf Figure 7.}  { Rest-frame $B$ absolute magnitude ($M_B$)
vs. logarithm of half-light radius ($r_{1/2}$) in kpc for our
confirmed sample of high redshift galaxies (open hexagons, large for
total objects and small for subclumps) as well as for a representative
sample of various types of local galaxies: ellipticals, dwarf
ellipticals/spheroidals, and spiral bulges (Bender \etal\ 1992);
spirals (total galaxy) and irregulars (de Vaucouleurs \etal\ 1991);
H~II galaxies (Telles 1995); and Compact Narrow Emission Line Galaxies
(CNELGs; Koo \etal\ 1994).  Despite their high luminosities
(i.e. $\sim$1-2 magnitudes above $L_*$), high redshift galaxies have
very small total sizes ranging from $1.7 < r_{1/2} < 7\ h_{50}^{-1}$
kpc (for $q_0 = 0.05$) with a median value $r_{1/2}=3.6 h_{50}^{-1}$
kpc.  For $q_0=0.5$, values would be $\sim$ 0.6 times those shown
here.  Radii have not been corrected for the WFPC2 PSF FWHM$\sim$
0\decsec14; such a correction will make the median half-light radius
10\% smaller, but will have a much larger effect on the smallest
sources.  This places them at the bright end of the sequence defined
by local HII galaxies, somewhat smaller than average massive galaxies
today but consistent with the sizes of spiral bulges.}






\begin{figure}
\vspace*{8in}
\includegraphics{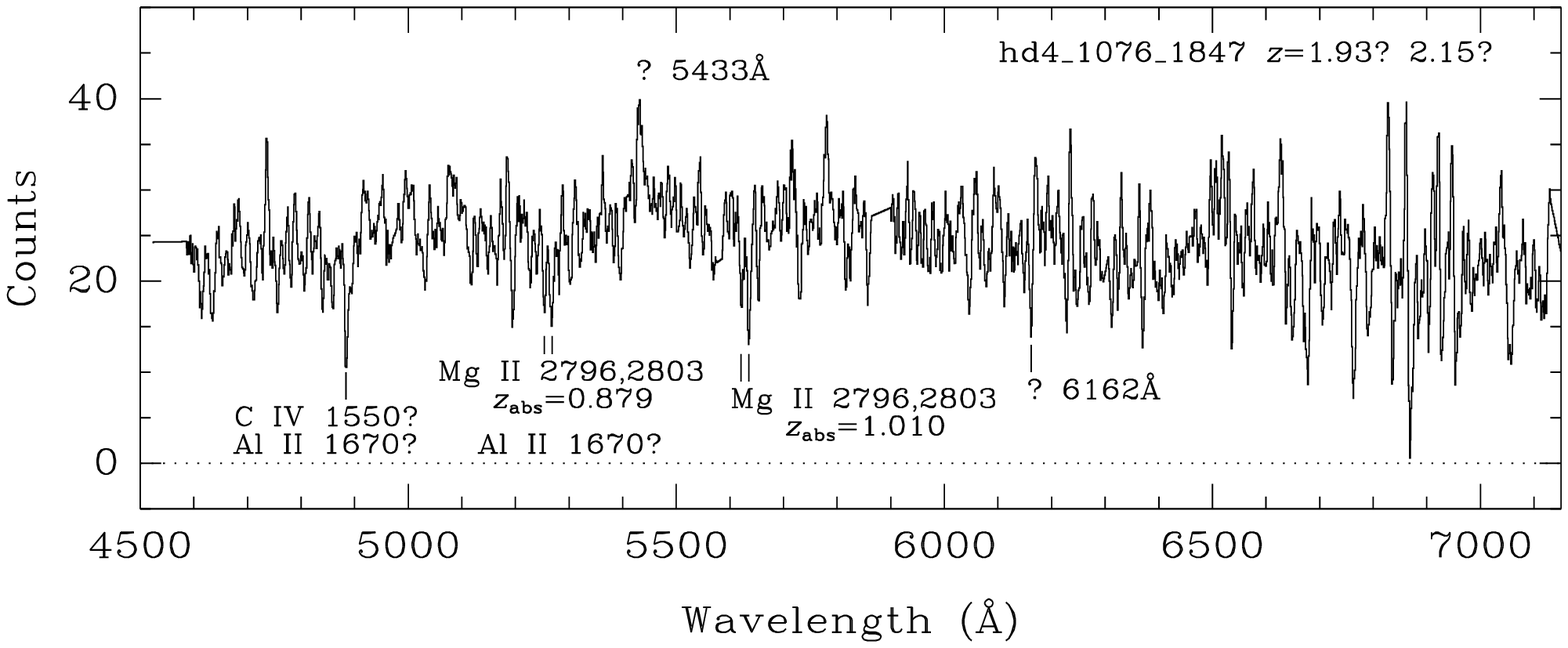}
\noindent {\bf Fig. 3}
\end{figure}

\begin{figure}
\vspace*{8in}
\includegraphics{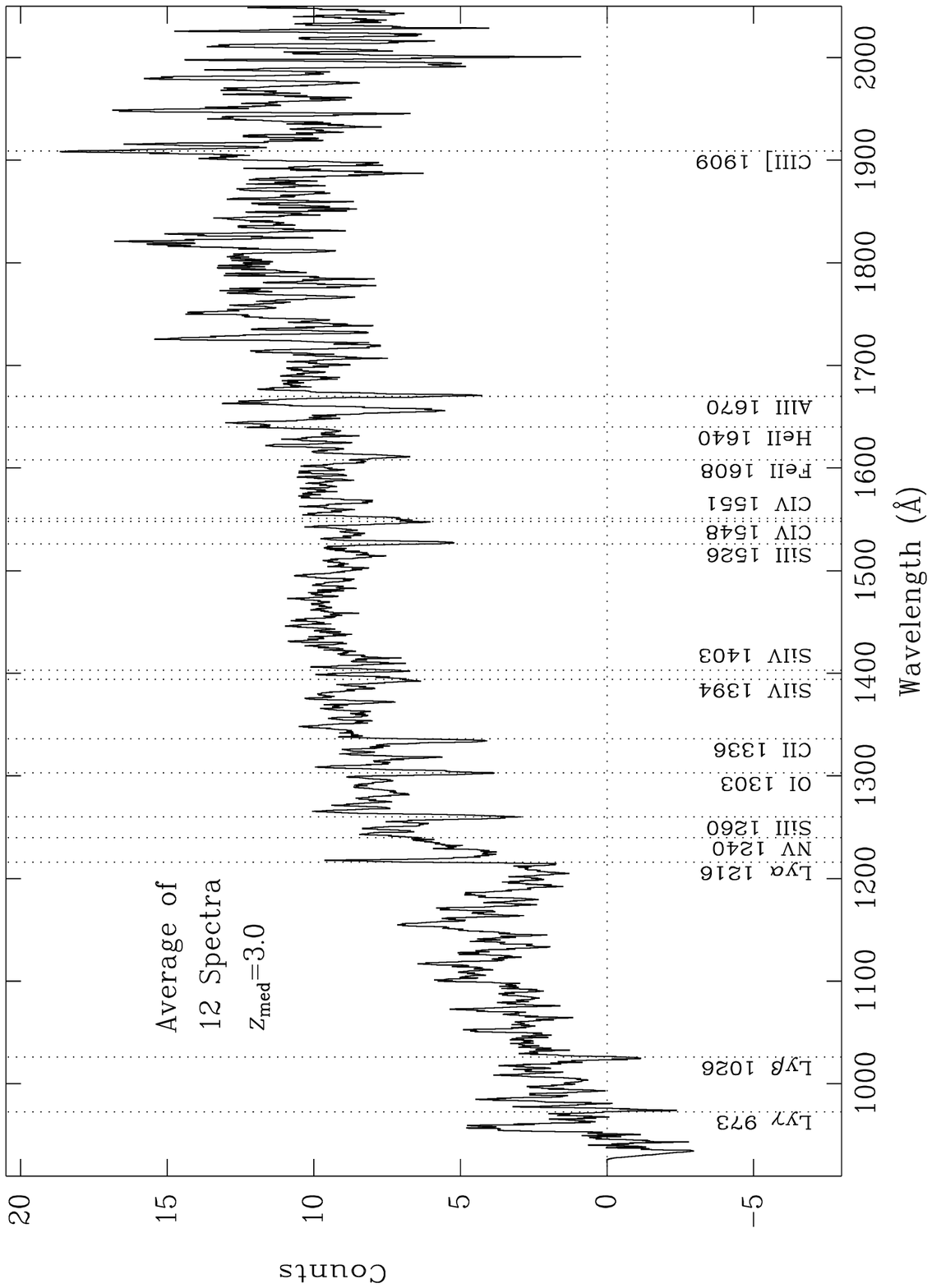}
\noindent {\bf Fig. 4}
\end{figure}

\begin{figure}
\vspace*{8in}
\includegraphics{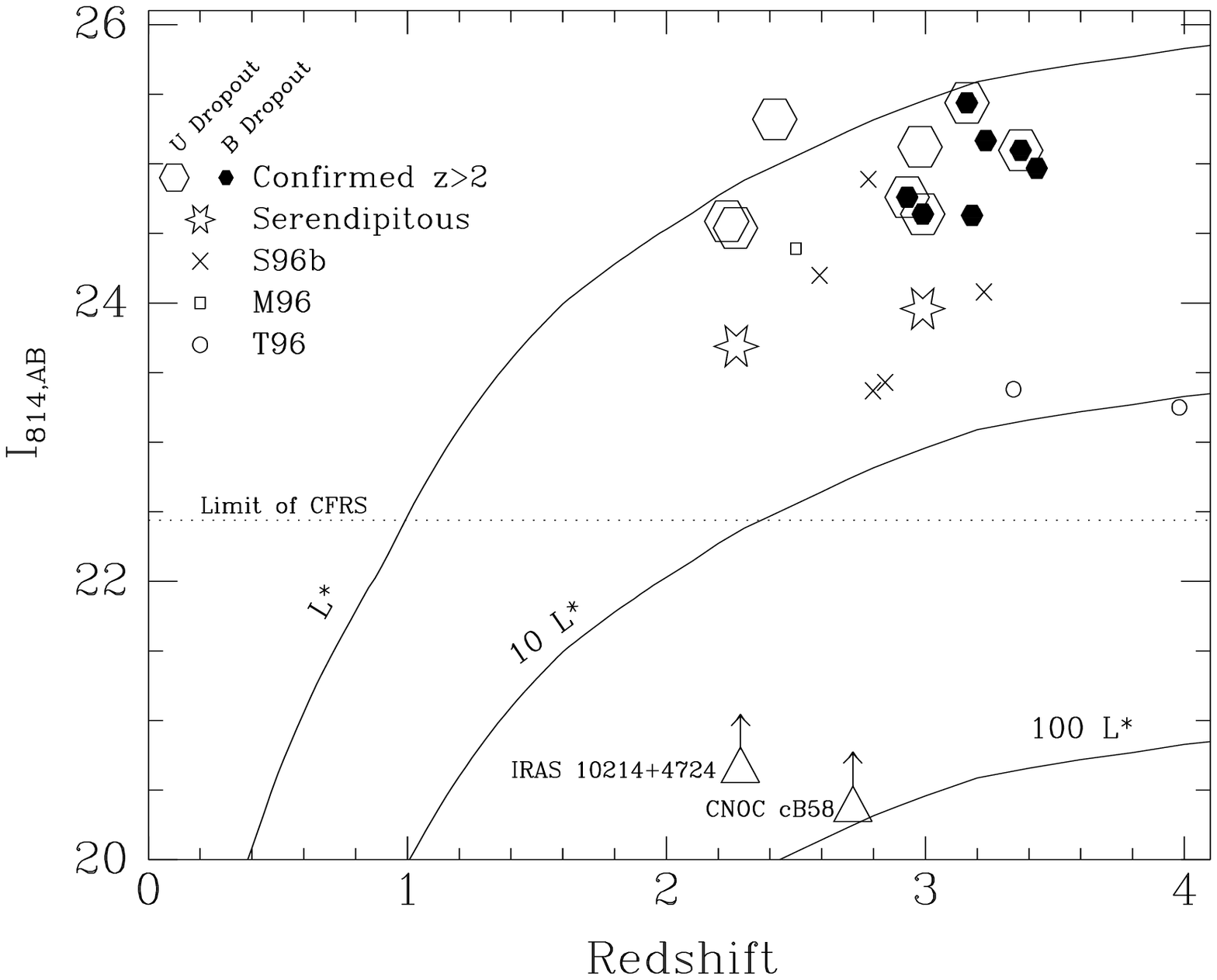}
\noindent {\bf Fig. 5}
\end{figure}

\begin{figure}
\vspace*{8in}
\includegraphics{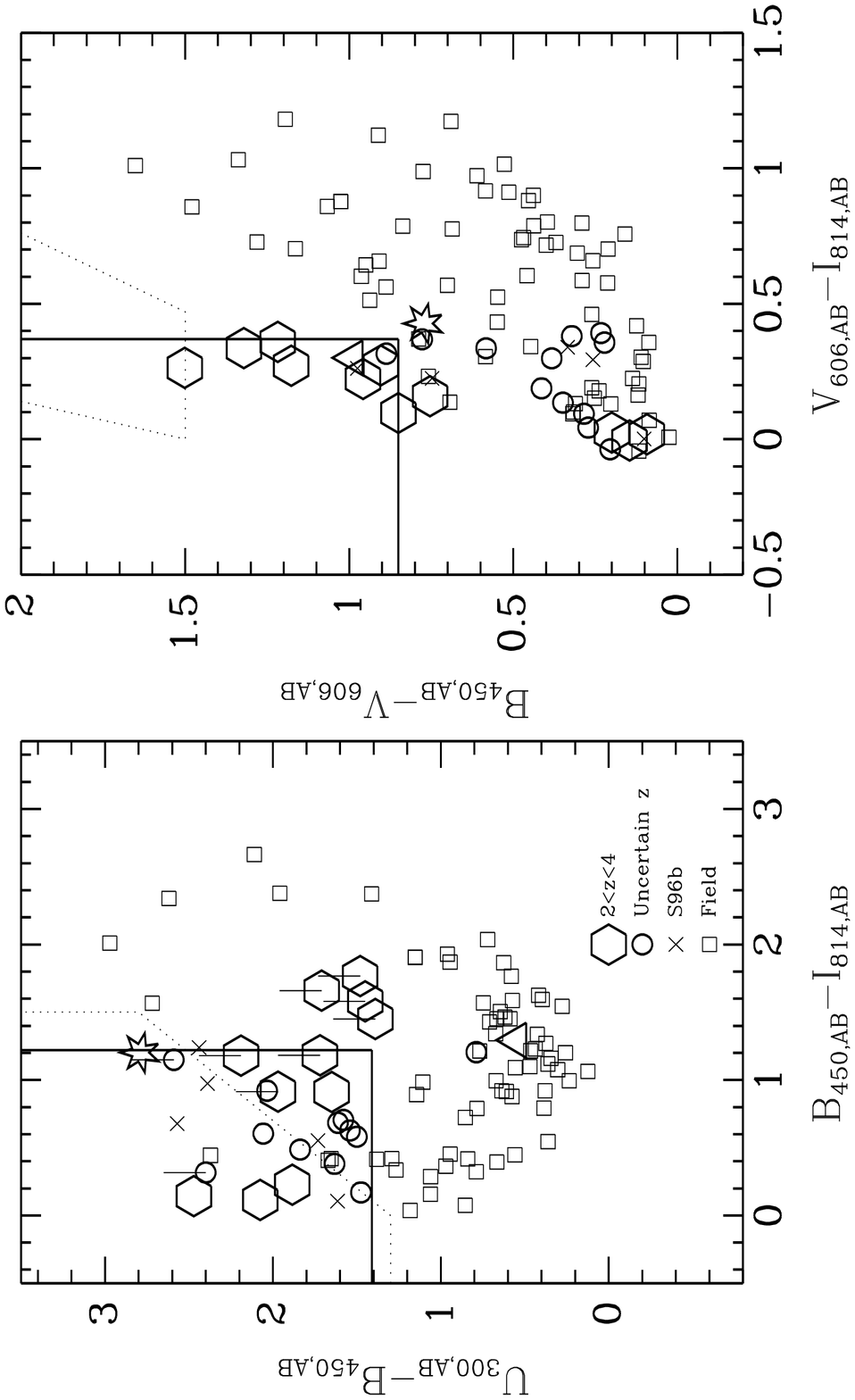}
\noindent {\bf Fig. 6}
\end{figure}

\begin{figure}
\vspace*{8in}
\includegraphics{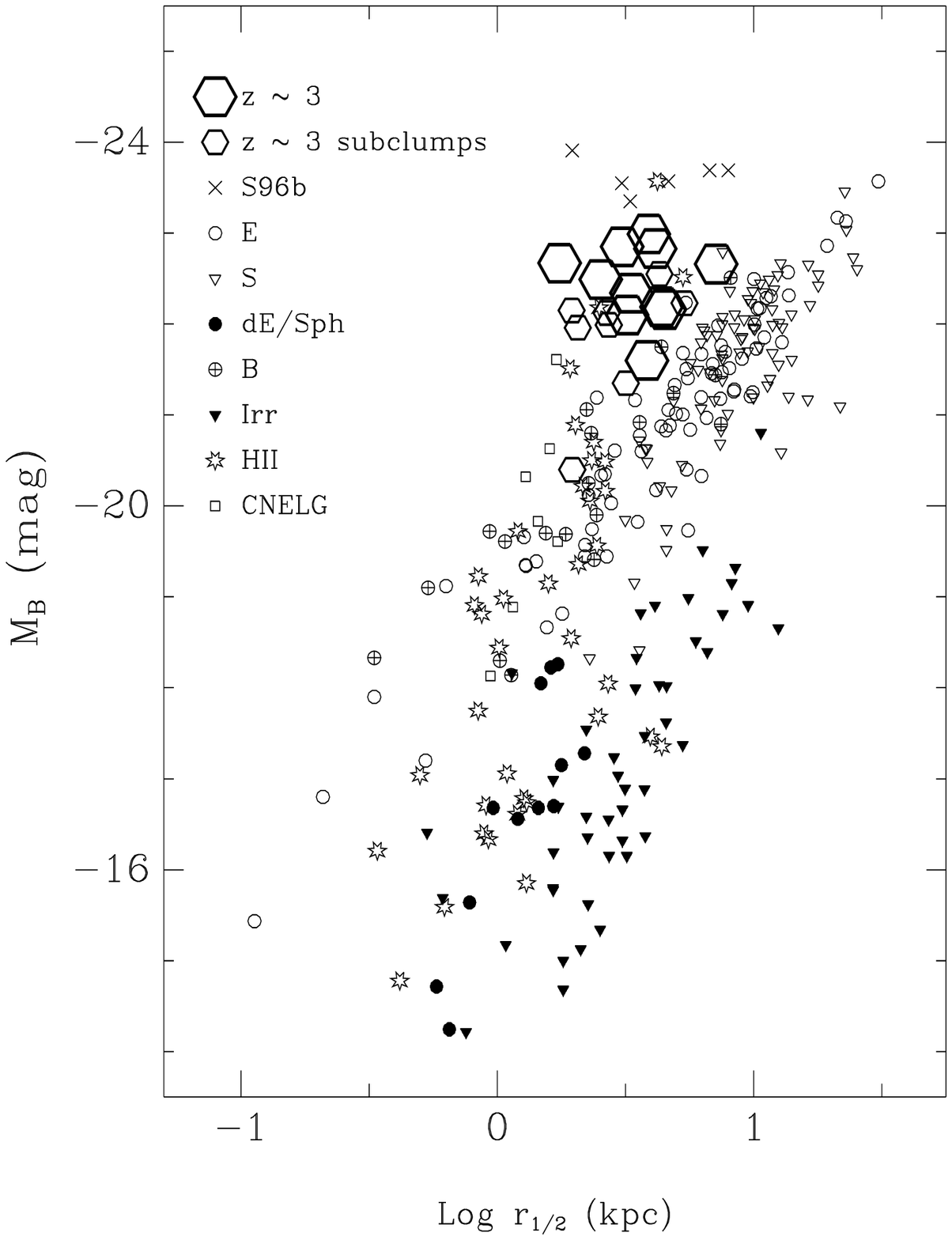}
\noindent {\bf Fig. 7}
\end{figure}

\end{document}